\documentclass[11pt]{article}
\usepackage{amsmath,amssymb,color,epsfig,slashed,cite}
\usepackage{graphicx}
\usepackage{setspace}

\usepackage{mathrsfs}

\usepackage{amsfonts}
\newcommand{\hoch}[1]{$\, ^{#1}$}

\makeatletter
\@addtoreset{equation}{section}
\makeatother

\newcommand{\be}{\begin{equation}}
\newcommand{\ee}{\end{equation}}
\newcommand{\bea}{\setlength\arraycolsep{2pt} \begin{eqnarray}}
\newcommand{\eea}{\end{eqnarray}}

\newcommand{\half}{{\textstyle{\frac{1}{2}}}}

\def\ndelta{\delta\hspace{-0.50em}\slash\hspace{-0.05em} }

\def\0{{\sst{(0)}}}
\def\1{{\sst{(1)}}}
\def\2{{\sst{(2)}}}
\def\3{{\sst{(3)}}}
\def\4{{\sst{(4)}}}
\def\5{{\sst{(5)}}}
\def\6{{\sst{(6)}}}
\def\7{{\sst{(7)}}}
\def\8{{\sst{(8)}}}
\def\sst#1{{\scriptscriptstyle #1}}

\def\scri{\mathscr{I}}

\def\td{\tilde}

\def\swedge{{\, \scriptstyle \wedge \,}}

\allowdisplaybreaks

\begin{document}

\begin{flushright}
\hfill{\hfill{}}

\end{flushright}

\vspace{15pt}
\begin{center}
{\Large {\bf Hamiltonian derivation of dual gravitational charges}}

\vspace{15pt}
{\bf Hadi Godazgar\hoch{1}, Mahdi Godazgar\hoch{2} and 
Malcolm J.\ Perry\hoch{3,4,5}}

\vspace{10pt}

\hoch{1} {\it Max-Planck-Institut f\"ur Gravitationsphysik (Albert-Einstein-Institut), \\
M\"ühlenberg 1, D-14476 Potsdam, Germany.}

\vspace{10pt}

\hoch{2} {\it School of Mathematical Sciences,
Queen Mary University of London, \\
Mile End Road, E1 4NS. United Kingdom.}

\vspace{10pt}

\hoch{3} {\it DAMTP, Centre for Mathematical Sciences,\\
 Cambridge University, Wilberforce Road, Cambridge CB3 OWA. United Kingdom.}

\vspace{10pt}

\hoch{4}{\it Department of Physics, Queen Mary University of London,\\
Mile End Road, E1 4NS. United Kingdom.}

\vspace{10pt}

\hoch{5}{\it Trinity College, Cambridge, CB2 1TQ. United Kingdom.}

 \vspace{15pt}
 
September 11, 2020

\vspace{20pt}

\underline{ABSTRACT}
\end{center}

\noindent We provide a Hamiltonian derivation of recently discovered dual BMS charges.  In order to do so, we work in the first order formalism and add to the usual Palatini action, the Holst term, which does not contribute to the equations of motion.  We give a method for finding the leading order integrable dual charges \textit{\`a la} Wald-Zoupas and construct the corresponding charge algebra.  We argue that in the presence of fermions, the relevant term that leads to dual charges is the topological Nieh-Yan term.  

\thispagestyle{empty}

\vfill
{\begin{center} hadi.godazgar@aei.mpg.de, m.godazgar@qmul.ac.uk, malcolm@damtp.cam.ac.uk \end{center}}

\pagebreak

\section{Introduction}

The intimate relation between symmetries and charges, as manifested in the Noether theorem, is a fundamental result of mathematical physics.  The application of these ideas in a gravitational setting is intricate, yet fundamental to almost any investigation involving gravity, from gravitational wave astrophysics to quantum gravity.  In this paper, we apply the prescription set out in Ref.\ \cite{letter}, which uses the covariant phase space formalism \cite{peierls,berg,CWitten,Crknovic88,Lw, Wald93, IW, WZ} to propose a systematic method for determining, in principle, all possible gravitational charges, to give a Hamiltonian derivation of a recently discovered tower of dual BMS charges \cite{dual0, dualex}.  One can think of dual BMS charges as generalisations of the Taub-NUT charge \cite{taub, nut, Ramaswamy, AS} in the same way that standard BMS charges \cite{Penrose, Dray:1984rfa, IW, WZ, BB, BarTro} generalise the notion of the Bondi linear four-momentum \cite{bondi, sachs}.

The recent interest on asymptotic charges, see for example Refs.\ \cite{Freidel:2019ohg, Alessio:2019cae, Laddha:2019yaj, porrati, Bart:2019pno, Lu:2019jus, Ruzziconi, Barnich:2019vzx, Choi:2019sjs, Campiglia:2020qvc, Henneaux:2020nxi, Gera:2020fvo, Barnich:2020ciy, Freidel:2020xyx, Freidel:2020svx}, is primarily motivated by the discovery of the importance of such charges in studies of gravitational scattering \cite{Strom:soft1, Strom:soft2, Strominger:2014pwa, Strom:lec} and the application of such ideas to black hole physics \cite{Hawking:2016msc, Hawking:2016sgy, Haco:2018ske}.  The potential success of such investigations and applications of asymptotic gravitational charges relies crucially on a good understanding of just how many asymptotic charges there are, and preferably a classification of all such charges, as envisaged in Ref.\ \cite{letter}.  The fact that in the last couple of years, two generalisations of asymptotic gravitational charges have been found \cite{fakenews, dual0, dualex} (see also \cite{conde}) indicates that there remains still much to be understood.  The fact that the dual BMS charges proposed in Refs.\ \cite{dual0, dualex} do not appear in previous analyses of BMS charges, such as Refs.\ \cite{IW, BB}, is particularly intriguing.  While it has been shown \cite{dualex} that the dual BMS charges satisfy the necessary properties of asymptotic charges and are therefore to be viewed as \textit{bona fide} charges, an \textit{ab initio} derivation has not been given.  This is the main aim of this paper: we apply the general formalism set out in Ref.\ \cite{letter} to provide a Hamiltonian derivation of the asymptotic dual BMS charges discovered in Refs.\ \cite{dual0, dualex}.

Previous classifications of asymptotic gravitational charges have, rather naturally, began with the Einstein-Hilbert action.  However, in Ref.\ \cite{letter}, it is argued that an investigation of asymptotic charges that solely focuses on the Einstein-Hilbert term will preclude other possible charges, such as dual charges.  One must entertain the existence of \emph{all} terms in the action whose equations of motion correspond to the Einstein equation, including the addition of terms that contribute trivially to the equations of motion.  The fact that different actions that give rise to the same equations of motion are fundamentally different in the quantum, or even semi-classical, theory is an old, and by now elementary, idea.  Indeed, such terms have been considered with a view to applications to the first law of black hole mechanics \cite{Jacobson:2015uqa} or to the study of particular solutions \cite{Araneda:2018orn}.  The inclusion of such terms whose addition do not change the equations of motion generally necessitates working in the first order formalism, which has been studied with a view to the definition of charges mainly in the context of the first law of black hole mechanics \cite{Frodden:2017qwh, DePaoli:2018erh, Oliveri:2019gvm, Aneesh:2020fcr} and in the context of asymptotic charges \cite{Barnich:2016rwk, Barnich:2019vzx, Barnich:2020ciy}.

In this paper, we concentrate on one such term, which one may add to the Einstein action without altering the Einstein equation, namely the Holst term \cite{Holst}.  We show that when added to the Palatini action (and more generally including other matter fields that do not give rise to torsion), the Holst term leads to dual gravitational charges.  In a setting, where there is non-trivial torsion, as a result, for example, of the existence of fermions, the Holst term is replaced by the topological Nieh-Yan term \cite{NiehYan}; see Refs. \cite{Mercuri_2008,Date:2008rb}.

The Holst term, or Nieh-Yan term in the presence of torsion, can, therefore, be viewed as the gravitational analogue of the $\theta$-term in electromagnetism. Note that in the latter case the application of the Noether theorem leads to magnetic charges and we show that an analogous picture holds in gravity.  

In the next section, \ref{sec:covrev}, we review the covariant phase space formalism and apply it in section \ref{sec:PH} to the Palatini-Holst theory. In order to make a link with standard and dual BMS charges, in section \ref{sec:asymp}, we state the boundary conditions that are of interest and derive the improper gauge transformations. The improper diffeomorphisms are given by the standard BMS generators, while we derive the large local Lorentz transformations\footnote{We use ``large'' instead of ``improper'' to avoid confusion with Lorentz transformations that include spatial reflections or time-reversal. The Lorentz transformations that we consider are proper in the latter sense.}. In section \ref{sec:charges}, we apply the covariant phase space analysis of section \ref{sec:PH} to these generators to find the asymptotic charges, showing that the Palatini action gives rise to the standard BMS charges, while the Holst term gives the dual charges. We apply the Wald-Zoupas method to find the integrable part of the leading order charges in section \ref{sec:int}.  In section \ref{sec:algebra}, we derive the charge algebra for leading order dual charges.  In the presence of torsion the Holst term needs to be modified, but we show in section \ref{sec:ferm} that we can nevertheless find dual charges in an asymptotically flat spacetime with torsion---this is achieved using the Nieh-Yan term. 

\section{Review of the covariant phase space formalism} \label{sec:covrev}

In this section, we review the covariant phase space formalism \cite{peierls,berg,CWitten,Crknovic88,Lw, Wald93, IW, WZ}, which provides a way of defining gravitational charges starting from a Lagrangian theory.  This section is based on the notation of Refs.\ \cite{Lw, Wald93, IW, WZ}.  

Given a top-form Lagrangian density $L$ for fields $\phi$, the Euler-Lagrange equations $E(\phi)$ are derived by varying the action,
\begin{equation}
 \delta L(\phi) = E(\phi) \delta \phi + d \theta(\phi, \delta \phi), \label{varL}
\end{equation}
where $\theta$, called the presymplectic potential\footnote{The reason why it is a presymplectic potential rather than a symplectic potential is that it is degenerate.  Indeed, the degenerate directions in phase space correspond to proper gauge transformations, i.e.\ those diffeomorphisms that vanish on the boundary.  In principle, we would need to factor out the degenerate subspaces in order to construct a true (or reduced) phase space.  However, in the covariant phase space formalism one works with the presymplectic manifold, which we simply call the phase space, avoiding the complications of having to work in the reduced phase space, which is no longer covariant.}, corresponds to the boundary terms, which appear when integrating by parts in order to derive the equations of motion.  As is clear from its definition above, $\theta$ is a one-form on phase space.

The exterior derivative on phase space of the presymplectic potential gives rise to a presymplectic form $\omega$, a two-form on phase space
\begin{equation}
 \omega(\phi, \delta_1 \phi, \delta_2 \phi) =  \delta_1 \theta(\phi, \delta_2 \phi) - \delta_2 \theta(\phi, \delta_1 \phi).
\end{equation}
Recall, from e.g.\ Ref.\ \cite{arnold}, that what defines a Hamiltonian flow is the existence of a Hamiltonian vector field $T$ on phase space whose 1-form dual on phase space is exact, i.e.\ using some local coordinates $A, B, \ldots$ on phase space
\begin{equation} \label{ham}
 (d H_T)_A = \omega_{AB} T^B.
\end{equation}
The phase space scalar $H_T$ thus derived is called a \emph{Hamiltonian}\footnote{While, technically the appropriate term is a Hamiltonian or a moment map, we choose to follow the more standard nomenclature by using the term ``charge'' or ``asymptotic charge'' henceforth. \label{ft:charge}} of the motion; it is conjugate in phase space to the transformation defined by $T$.  In other words, the direction $T$ in phase space corresponds to an integral curve.  In canonical coordinates the above equation reduces to Hamilton's equations.  We translate the above expression to the covariant phase space language we have been using by noting that as a vector field on phase space, $T$ corresponds to a particular transformation of the fields.  Hence, equation \eqref{ham} is equivalent to
\begin{equation} \label{hamiltonian}
 \delta H_\tau = \int_\Sigma \omega(\phi, \delta \phi, \delta_\tau \phi),
\end{equation}
where $\tau$ is some transformation parameter and we integrate over some Cauchy surface $\Sigma.$  Thus, we have a charge associated with a transformation generated by $\tau$ if the right hand side of equation \eqref{hamiltonian} is integrable.  Moreover, it would be desirable to convert the integral to a boundary integral.  This is because, we will be primarily interested in asymptotic symmetry generators, i.e.\ solutions that have a specific asymptotic form and corresponding symmetry generators that keep this form intact.  For the asymptotic generators to define a \textit{bona fide} charge, it would make sense for it to be given in terms of a boundary integral.  This would be the case, were $\omega(\phi, \delta \phi, \delta_\tau \phi)$ an exact form in spacetime.  

For concreteness, let us consider diffeomorphisms generated by vector fields $\xi$.  In this case, $\delta_\xi$ corresponds to a Lie derivative so that
\begin{equation}
 \omega(\phi, \delta \phi, \mathcal{L}_\xi \phi) = \delta \theta(\phi, \mathcal{L}_\xi \phi) - \mathcal{L}_\xi \theta(\phi, \delta \phi).
\end{equation}
Using the Cartan magic formula
\begin{equation} \label{magic}
 \mathcal{L}_\xi = d \iota_\xi + \iota_\xi d,
\end{equation}
the second term
\begin{align}
 \mathcal{L}_\xi \theta(\phi, \delta \phi) &= d \iota_\xi \theta(\phi, \delta \phi) + \iota_\xi d \theta(\phi, \delta \phi) \notag \\
                                           &\approx d \iota_\xi \theta(\phi, \delta \phi) + \iota_\xi \delta L(\phi),
\end{align}
where we have used equation \eqref{varL} and $\approx$ denotes an expression that is valid on-shell for the field, as well as its variation.  Therefore,
\begin{equation}
 \omega(\phi, \delta \phi, \mathcal{L}_\xi \phi) = \delta \left[\theta(\phi, \mathcal{L}_\xi \phi) -\iota_\xi L(\phi) \right] - d \iota_\xi \theta(\phi, \delta \phi).
\end{equation}
The expression in the square brackets above is called a Noether current $j$ and one can show that it is closed: consider the exterior derivative of the Noether current
\begin{align}
 d j_\xi &\equiv d \left[\theta(\phi, \mathcal{L}_\xi \phi) -\iota_\xi L(\phi) \right] \notag \\
     &= d \theta(\phi, \mathcal{L}_\xi \phi) - (\mathcal{L}_\xi - \iota_\xi d) L(\phi),
\end{align}
where we have again used the magic formula \eqref{magic}.  Now, using the fact that $L$ is a top-form so that $dL=0$ and equation \eqref{varL}, we find that
\begin{equation}
 dj_\xi \approx 0.
\end{equation}
The Poincar\'e lemma implies that \cite{Wald90, IW2}
\begin{equation} \label{current}
 j_\xi = dQ_\xi = \theta(\phi, \mathcal{L}_\xi \phi) -\iota_\xi L(\phi),
\end{equation}
where $Q_\xi$ is called the Noether charge.  This means that
\begin{equation}
 \omega(\phi, \delta \phi, \mathcal{L}_\xi \phi) \approx d \left[\delta Q_\xi - \iota_\xi \theta(\phi, \delta \phi)\right]
\end{equation}
so that
\begin{equation} \label{mm}
 \delta H_\xi = \int_{\partial \Sigma} \Big\{ \delta Q_\xi - \iota_\xi \theta(\phi, \delta \phi) \Big\},
\end{equation}
where the integral is a surface integral over a cross-section $\partial \Sigma$ of ``infinity''---we will make this more precise in section \ref{sec:asymp}.

What remains to consider is whether the charge exists at all, i.e.\ whether equation \eqref{mm} is integrable \cite{WZ}.  Certainly, a necessary (and sufficient \cite{WZ}) condition is that 
\begin{equation}
 (\delta_1 \delta_2 - \delta_2 \delta_1) H_\xi = - \int_{\partial \Sigma} \iota_\xi \omega(\phi, \delta_1 \phi, \delta_2 \phi) = 0,
\end{equation}
which is not generically satisfied.  This obstruction to the existence of a charge is directly related to the existence of flux at infinity and is resolved by taking the flux into account \cite{WZ}.  In order to make it clear that the expression in equation \eqref{mm} is not necessarily integrable, following Ref.\ \cite{BarTro} we rewrite equation \eqref{mm} as
\begin{equation} \label{diffQ}
 \ndelta H_\xi = \int_{\partial \Sigma} \Big\{ \delta Q_\xi - \iota_\xi \theta(\phi, \delta \phi) \Big\}.
\end{equation}
Clearly, we can rewrite the above equation as
\begin{equation} \label{Q:intnint}
  \ndelta H_\xi = \delta \mathcal{H}_\xi + \mathcal{N}_\xi,
\end{equation}
i.e.\ we can split the expression in terms of an integrable part given by the true variation of an integrable charge $\mathcal{H}_\xi$ and a non-integrable part, whose existence is directly related to the existence of flux at infinity.  However, the splitting above is ambiguous:
\begin{equation} \label{Q:amb}
 \mathcal{H}_\xi \rightarrow \mathcal{H}_\xi + \mathcal{I}, \qquad \mathcal{N}_\xi \rightarrow \mathcal{N}_\xi - \delta \mathcal{I}.
\end{equation}
Ref.\ \cite{WZ} gives a prescription for fixing this ambiguity based on reasonable criteria such as the fact that $\mathcal{N}_\xi$ be locally constructed from dynamical fields and their derivatives and that it vanish in the case where there is no radiation.  Based on these criteria Wald-Zoupas \cite{WZ} propose that
\begin{equation} \label{nonint}
 \mathcal{N}_\xi = - \int_{\partial \Sigma} \iota_\xi \Theta(\phi, \delta \phi),
\end{equation}
where $\Theta$ is the potential for the pull-back of the presymplectic 2-form to infinity $\bar{\omega}$
\begin{equation} \label{bigTh}
 \bar{\omega}(\phi, \delta_1 \phi, \delta_2 \phi) =  \delta_1 \Theta(\phi, \delta_2 \phi) - \delta_2 \Theta(\phi, \delta_1 \phi).
\end{equation}
Hence, the integrable charge is given by
\begin{equation} \label{intQ}
 \delta \mathcal{H}_\xi = \int_{\partial \Sigma} \delta Q_\xi - \iota_\xi \theta(\phi, \delta \phi) + \int_{\partial \Sigma} \iota_\xi \Theta(\phi, \delta \phi).
\end{equation}

In Einstein gravity given by the Einstein-Hilbert action, these charges are precisely the BMS charges in the context of asymptotically flat boundary conditions. The goal in the next sections is to apply this formalism to first order actions. 

\section{Gravitational theory in first order formalism}
\label{sec:PH}

We consider as the gravitational action the Palatini action, which is a first order tetrad formulation of Einstein's theory plus the Holst term \cite{Holst}. As noted in the introduction, General relativity in the first order formalism, with the Holst term and without, has already been considered in the literature principally in the context of the first law of black hole mechanics. Indeed much of the covariant phase space analysis of this system has already been studied in \cite{DePaoli:2018erh, Oliveri:2019gvm}; we revisit the covariant phase space analysis of the Palatini-Holst theory and identify new gravitational charges, namely dual charges \cite{dual0, dualex}.

The action that we consider is
\begin{equation}
 S_{PH} = \frac{1}{16 \pi G} \int_{\mathcal M} P_{abcd} \, \mathcal{R}^{ab}(\omega) \swedge e^{c} \swedge e^{d},
 \label{EHH}
\end{equation}
where Latin indices $a,b,c, \dots$ denote tangent space indices, $e^a$ is the vierbein and $\omega$ is the spin connection and is treated as an independent field. We denote the fields collectively as $\phi= \{ e, \omega\}.$  The 2-form Riemann curvature
\begin{equation}
 \mathcal{R}^{a}{}_{b}(\omega) = d \omega^{a}{}_{b} + \omega^{a}{}_{c} \swedge \omega^{c}{}_{b}
\end{equation}
and the tensor 
\begin{equation} \label{P}
 P_{abcd} = \frac12 \varepsilon_{abcd} + i \, \lambda \, \eta_{a [c} \eta_{d] b},
\end{equation}
where in our convention the antisymmetrisations have weight 1 and $\eta$ is the flat space metric.  

The parameter $\lambda$ is inversely proportional to the Barbero-Immirzi parameter in loop quantum gravity (see \cite{DePaoli:2018erh} and references therein). In our case, we will consider it to be a general parameter.
When $\lambda = 0,$ this action is the Palatini action, while the term proportional to $\lambda$ is the Holst term. It is worth noting that if the spin connection is viewed as depending on the vierbein and solving Cartan's first structure equation with vanishing torsion
\begin{equation}
 d e^a + \omega^a{}_{b} \swedge e^b=0,
 \label{Cartan1}
\end{equation}
the Holst term becomes trivial as a result of the algebraic Bianchi identity. However, in the first order formalism, where $e$ and $\omega$ are treated as independent fields, the above argument does not apply; hence the Holst term is non-trivial. Of course, as we shall show below, the Holst term is on-shell zero, but this is no different to the fact that the Palatini term vanishes on-shell by virtue of the Einstein equation. 

The tensor $P$ is invertible, as a $6 \times 6$ tensor $P_{[ab][cd]}$, where we think of the first and last two antisymmetric indices as a single bivector index, when $\lambda \neq \pm 1$, with inverse
\begin{equation}
 P^{-1}_{abcd} = \frac{1}{2 (\lambda^2 - 1)} \left( \varepsilon_{abcd} - 2 \, i \, \lambda \eta_{a [c} \eta_{d] b} \right).
 \label{Pinv}
\end{equation}

When $P$ is invertible, the variation of the action \eqref{EHH} with respect to the spin connection gives rise to the torsion-free condition \eqref{Cartan1}, while the variation of the vierbein gives the vacuum Einstein equation, \emph{viz}.\ Ricci flatness.  Therefore, the addition of the Holst term has not materially affected the theory, at least at the level of the equations of motion.  However, the inclusion of the Holst term does significantly affect the Hamiltonian analysis of the theory and the symplectic current therefrom.  It is this difference that allows a derivation of dual gravitational charges starting from an action.  Therefore, any treatment of a gravitational system that takes dual charges seriously must also take the Holst term seriously.  

Inspecting action \eqref{EHH}, it is straightforward to see that the presymplectic potential  is 
\begin{equation} \label{EHH:sympot}
 \theta(\phi, \delta \phi) = \frac{1}{16 \pi G} P_{abcd} \, e^a\swedge e^b \swedge \delta \omega^{cd}.
\end{equation}
Note that the presymplectic potential does not depend on $\delta e.$

Before we study the set of charges that can be derived from a covariant phase space analysis of this theory, we need to define the class of solutions we are interested in.  This will give us the set of transformations that lead to the existence of non-trivial charges.  Therefore, we turn now to the definition of asymptotically flat spacetimes and an analysis of their asymptotic symmetry generators, which allows us to find the associated charges or moment maps.

\section{Asymptotic flatness and symmetries}
\label{sec:asymp}

We consider asymptotically flat spacetimes $\mathcal{M}$  as a triplet $(\mathcal{M} \cup \scri, e, \omega),$  with boundary conditions on the fields, the vierbein and spin connection, at null infinity $\scri$ such that the relevant quantities are well-defined at $\scri$. The space $\mathcal{M} \cup \scri$ is the unphysical space corresponding to the conformal compactification of $\mathcal{M}$.\footnote{In this paper, we will consider future null infinity $\scri^+$, but the same methods can easily be adapted to past null infinity as well.}  In fact, we will not explicitly compactify and instead follow the Bondi-Sachs approach \cite{bondi, sachs}, albeit in a tetrad form, as explained below.

\subsection{Boundary conditions}

The vierbein $e_{\mu}^{a}$ has Greek spacetime indices $\mu, \nu, \dots$ and tangent space indices denoted by Latin letters $a, b, \dots$. Tangent space indices are lowered and raised using the flat metric (and its inverse),~\footnote{This form of the flat metric requires a complex basis of zweibeine for the two-sphere cross-sections of $\scri$. However, in  practice we do not choose a particular basis for the 2-space and all of our expressions are covariant along the 2-sphere directions.}
\begin{equation}
 \eta = \begin{pmatrix}
         0 & -1 & 0 & 0 \\
         -1 & 0 & 0 & 0 \\
         0 & 0 & 0&1 \\
         0 & 0 & 1& 0
        \end{pmatrix}.
 \end{equation}

In components, where the coordinates $X^\mu=(u,r,x^I)$, the (inverse) vierbein is given by 
\begin{align}
 e^0 &= \frac12 F du + dr, \hspace{23mm} e_{0} = \partial_r \notag \\
 e^1 &= e^{2 \beta} \, du, \hspace{31mm} e_1 = e^{- 2 \beta} \left( \partial_u - \frac12 \, F \, \partial_r + C^I \, \partial_I \right), \label{vielbeine} \\
 e^i &= r \,  E^{i}_{I} \left( dx^I - C^I du \right), \hspace{10mm} e_{i} = \frac{1}{r} \, E_{i}^{I} \, \partial_I, \notag 
\end{align}
where $I,J, \dots $ denote coordinates on a 2-sphere, e.g.\ $x^I= (\theta, \phi)$, and we denote tangent space indices on the 2-sphere with indices $i,j, \dots\; $.

The boundary conditions for the fields can now be given in terms of the components above,
\begin{align}
 F(u, r, x^I) &= 1+ \frac{F_0(u, x^I)}{r} + o(r^{-1}), \hspace{4mm} \beta(u, r, x^I) = \frac{\beta_0(u, x^I) }{r^2} + o(r^{-2}), \notag \\
 C^I(u, r, x^I) &= \frac{C_{0}^I(u,x^I)}{r^2} + o(r^{-2}), \hspace{7mm}  E^{i}_{I} (u, r, x^I)= \hat{E}^{i}_{I}(x^I) + \frac{C_{IJ} \hat{E}^{iJ}}{2r} + o(r^{-1}),   
 \label{falloffs}
\end{align}
where $C_{IJ}$ is a trace-free, symmetric tensor and $\hat{E}$ is the zweibein on a round sphere, i.e.
\begin{equation}
 \gamma_{IJ} = \hat{E}^i_I \, \hat{E}^j_J \; \eta_{ij}
\end{equation}
with $\gamma_{IJ}$ the metric on the round 2-sphere.  Note that $\hat{E}^{iJ} = \gamma^{IJ} \hat{E}^i_I$.  Unless explicitly stated, throughout this paper, $I,J,\ldots$ indices on tensors defined on the 2-sphere are lowered and raised using only $\gamma_{IJ}$ and its inverse.  Furthermore, we require that 
\begin{equation} \label{con:det}
 \det E_I^i = \det \hat{E}_I^i
\end{equation}
so that in $(\theta, \phi)$ coordinates
\begin{equation}
 \det E_I^i = \sin \theta.
\end{equation}

These boundary conditions imply the weakest boundary conditions on the metric in order to have well-defined quantities at $\scri$, namely they are equivalent at leading order to the boundary condition used by, for example, Sachs \cite{sachs}.

The torsion-free (on-shell) spin connection is given by the vielbein postulate
\begin{equation}
 \nabla_\mu e^a_\nu = \partial_\mu e^a_\nu - \Gamma^{\rho}_{\mu \nu} e^a_\rho + \omega_{\mu}{}^{a}{}_{b} e^b_\nu = 0;
\end{equation}
hence
\begin{equation}
 \omega_{\mu}{}^{a}{}_{b}= e^{\nu}_{b} \left( \Gamma_{\mu \nu}^{\rho} e^{a}_{\rho} - \partial_{\mu} e_{\nu}^{a} \right),
\end{equation}
where  $\Gamma$ is the affine connection, which coincides with the Christoffel symbols as a result of vanishing torsion.  Using this fact, the spin connection can also be written as
\begin{equation}
 \omega_{\mu \, ab} = e^{\rho}_{[a} e^{\sigma}_{b]} \left( e_{\sigma \, c} \, \partial_{\mu} e_{\rho}^{c} + \partial_{\sigma} g_{\rho \mu} \right), \label{omega}
\end{equation}
where
\begin{equation}
 g_{\mu \nu} = e_{\mu}^{a} \, e_{\nu}^{b} \, \eta_{ab}
\end{equation}
and $e_{\sigma \, c} = g_{\sigma \tau} e^\tau_c = \eta_{cd} e_\sigma^d$.  We list the metric, inverse metric and spin connection components associated with vierbein \eqref{vielbeine} in appendix \ref{app:spin}.

\subsection{Asymptotic symmetry generators} 

We find the diffeomorphisms and Lorentz transformations that preserve the boundary conditions presented in the previous section. 

The transformation of the inverse vierbein is 
\begin{equation}
 \delta e^{\mu}_{a} = \xi^{\nu} \partial_{\nu} e^{\mu}_{a} - e^{\nu}_a \partial_{\nu} \xi^{\mu} + \Lambda_{a}{}^{b} e_{b}^{\mu}. \label{vp}
\end{equation}
The boundary conditions on the vierbein, \eqref{vielbeine} and \eqref{falloffs}, are preserved for diffeomorphisms of the form 
\begin{gather}
 \xi^{u} = f(u, x^I)= s(x^I) + \frac{u}{2} \, D_{I} Y^I,   \qquad 
\xi^{r} = \frac{r}{2} \left( C^I \partial_{I} f - D_{I} \xi^I \right),  \notag \\[3pt]
 \xi^{I} = Y^{I} - \int_{r}^{\infty} dr' \, \frac{e^{2 \beta}}{r'^2} \, h^{IJ} \partial_{J} f, \label{xi}
\end{gather}
where $D$ is the covariant derivative on the round sphere,
\begin{equation}
 h^{IJ} = E^{I}_{i} E^{J}_{j} \eta^{ij}
\end{equation}
and $Y^I(x^I)$ are conformal Killing vectors on the sphere\footnote{As emphasised before, unless stated otherwise, we always lower/raise $I,J, \ldots$ indices on tensors defined on the 2-sphere only with the metric on the round 2-sphere $\gamma_{IJ}$ and its inverse.}
\begin{equation} \label{Y:CKV}
 D_{(I} Y_{J)} = \half\, D_K Y^K \, \gamma_{IJ}.
\end{equation}
These are the familiar BMS transformations~\cite{bondi}. 
And the Lorentz transformations that preserve the boundary conditions are 
\begin{align}
 \Lambda_{01} &= - \partial_{r} \xi^r, \hspace{34mm} \Lambda_{0 i} = \frac{e^{2 \beta}}{r} \, E_{i}^{I} \partial_{I} \xi^u, \notag \\[3pt]
 \Lambda_{1 i} &=  \frac{E_{i}^{I}}{2 r} \left( F \partial_{I} \xi^u + 2 \, \partial_{I} \xi^{r}\right), \hspace{10mm} \Lambda_{ij} = \gamma_{IJ} \hat{E}^I_{[i} \mathcal{L}_{Y} \hat{E}_{j]}^{J} + o(r^0). \label{lorentz}
\end{align}

One can show that the BMS generators satisfy the following identities
\begin{align}
 \nabla_{r} \xi^{u} &= 0, \label{drxiu} \\
 g_{a(r}\nabla_{I)} \xi^{a} &= 0, \label{symrI} \\
 \nabla_{I} \xi^{I} &= C^{I} \nabla_{I} \xi^{u}. \label{divxiI}
\end{align}

\section{Asymptotic charges} \label{sec:charges}

The gauge transformations of the theory \eqref{EHH} are diffeomorphisms and local Lorentz transformations, with the asymptotic symmetry transformations given by the improper coordinate transformations generated by the vector fields given in equation \eqref{xi}, BMS transformations, and large Lorentz transformations with parameters given in equation \eqref{lorentz}---these are local Lorentz versions of BMS transformations. The question that we address in this section is what are the asymptotic charges corresponding to these improper gauge transformations.  We consider diffeomorphisms and Lorentz transformations in turn.  However, it should be emphasised that strictly diffeomorphisms and Lorentz transformations ought to be considered together, since the asymptotic symmetry transformations are constructed from the simultaneous action of diffeomorphisms and Lorentz transformations.\footnote{One could equally derive the asymptotic symmetries corresponding to the independent action of diffeomorphisms and Lorentz transformations.  However, the conditions in this case would be too strong and preclude the BMS group.}  It turns out that for the theories that we consider in this paper, there is a clean decoupling of the two sets of transformations, which allows them to be considered separately.\footnote{This is not the case, for example, for the Pontryagin and Gauss-Bonnet terms \cite{letter}.}  We choose to take advantage of this feature to consider them separately for ease of exposition.

\subsection{Diffeomorphisms: standard and dual BMS charges}

In section \ref{sec:covrev}, we reviewed how asymptotic diffeomorphism charges are defined and showed that
\begin{equation} \label{eq:diffq}
 \ndelta H_\xi \equiv \int_{\Sigma} \omega(\phi, \delta \phi, \mathcal{L}_\xi \phi) = \int_{\partial \Sigma} \Big\{ \delta Q_\xi - \iota_\xi \theta(\phi, \delta \phi) \Big\},
\end{equation}
where $\partial \Sigma$ is a cross-section of $\scri^+$ and
\begin{equation}
 dQ_\xi = \theta(\phi, \mathcal{L}_\xi \phi) -\iota_\xi L(\phi).
\end{equation}
Since the action \eqref{EHH} vanishes on-shell, the above equation reduces, on-shell, to
\begin{equation}
 d Q_{\xi} = \theta(\phi, \mathcal{L}_{\xi} \phi).
\end{equation} 
From equation \eqref{EHH:sympot},
\begin{equation} \label{EHH:sympotdiff}
 \theta(\phi, \mathcal{L}_{\xi} \phi) = \frac{1}{16 \pi G} P_{abcd} \, \mathcal{L}_{\xi} \omega^{ab} \swedge e^c \swedge e^d.
\end{equation}
Using the magic formula \eqref{magic}, it is simple to show that the Noether charge is
\begin{equation} \label{Noether:diff}
  Q_{\xi} = \frac{1}{16 \pi G} P_{abcd} \, \iota_{\xi}\omega^{ab} \,e^c\swedge e^d.
\end{equation}
Therefore, using equations \eqref{EHH:sympot} and \eqref{Noether:diff}, equation \eqref{eq:diffq} becomes
\begin{align}
 \ndelta H_{\xi} &= \frac{1}{16 \pi G} P_{abcd} \int_{\partial \Sigma} \,\left[ \delta \left( \iota_{\xi}\omega^{ab} \,e^c\swedge e^d \right) - \iota_{\xi}\left(\delta \omega^{ab} \swedge \,e^c\swedge e^d \right)\right] \notag \\[3mm]
 &= \frac{1}{8 \pi G} P_{abcd} \int_{\partial \Sigma} \,\left[ \iota_{\xi}\omega^{ab} \, \delta e^c + \iota_{\xi}e^c \, \delta \omega^{ab} \right]\swedge e^d.  
 \label{eq:diffq1}
\end{align}
Consider
\begin{equation}
 \delta e^{[c} \swedge e^{d]}\, |_{\partial \Sigma}.
\end{equation}
In components this would be equal to
\begin{equation} \label{vee}
2 \left(\delta e^{[c}_{[I} \right) e^{d]}_{J]} = \delta \left(e^{[c}_{[I}\, e^{d]}_{J]}\right) = r^2 \delta^{cd}_{ij} \delta \left(E^{i}_{[I}\, E^{j}_{J]}\right) 
= r^2 \delta^{cd}_{ij} \delta \left(\hat{E}^{i}_{[I}\, \hat{E}^{j}_{J]}\right) = 0,
\end{equation}
where in the second equality we use equations \eqref{vielbeine}, in the third equality we use equation \eqref{con:det} and in the final equality we use the fact that the variation of the zweibein on the round sphere is trivial.  Therefore, equation \eqref{eq:diffq1} reduces to
\begin{equation}
 \ndelta H_{\xi} = \frac{1}{8 \pi G} P_{abcd} \int_{\partial \Sigma} \, \iota_{\xi}e^c \, \delta \omega^{ab} \swedge e^d.
\end{equation}
Using equation \eqref{P}, we rewrite this expression as
\begin{equation} \label{nH}
 \ndelta H_{\xi} = \ndelta \mathcal{Q}_\xi + i\, \lambda \, \ndelta \td{\mathcal{Q}}_\xi,
\end{equation}
where
\begin{equation} \label{nHSD}
\ndelta \mathcal{Q}_\xi = \frac{1}{16 \pi G}\, \varepsilon_{abcd} \int_{\partial \Sigma} \, \iota_{\xi}e^c \, \delta \omega^{ab} \swedge e^d, \qquad
\ndelta \td{\mathcal{Q}}_\xi = \frac{1}{8 \pi G} \int_{\partial \Sigma} \, \iota_{\xi}e^a \, \delta \omega_{ab} \swedge e^b
\end{equation}
are to be viewed as the standard (``electric'') and dual (``magnetic'') BMS charges, respectively.  Now, we consider each of these expressions separately.

\subsubsection{Standard BMS charges}

The standard BMS charge is
\begin{equation}
 \ndelta \mathcal{Q}_\xi = \frac{1}{16 \pi G}\, \varepsilon_{abcd} \int_{\partial \Sigma} \, \iota_{\xi}e^c \, \delta \omega^{ab} \swedge e^d.
\end{equation}

Using equation \eqref{omega}, it can be shown that\footnote{A repeated use of the Schouten identity
\begin{equation}
 5 \varepsilon_{[\mu \nu \rho \sigma} X_{\tau]} = 2 \varepsilon_{\rho \sigma \tau [\mu} X_{\nu]} + 3 \varepsilon_{\mu \nu [\rho \sigma} X_{\tau]} = 0
\end{equation}
for an arbitrary $X$ is required.
}
\begin{align}
\ndelta \mathcal{Q}_\xi =  \frac{3}{32 \pi G} \int_{\partial \Sigma} \,  \varepsilon_{\mu \nu \rho \sigma} \, \left( \, g^{\eta [\tau} \xi^{\sigma} \nabla^{\rho]} \delta g_{\eta \tau} + \xi^{[\tau} \nabla_{\tau} \left( e^{\sigma}_{a} \delta e^{\rho] \, a} \right)  \right) dx^{\mu} \swedge dx^{\nu}.
\label{integrand}
\end{align} 
Of course, $\mu \nu = IJ$ in the expression above.  Let us consider the second term, 
\begin{align}
 3 \, \varepsilon_{\mu \nu \rho \sigma} \, \xi^{[\tau} \nabla_{\tau} \left( e^{\sigma}_{a} \delta e^{\rho] \, a} \right) =  -2 \, \nabla_{[\mu} \left( \varepsilon_{ \nu] \rho \sigma \tau} \xi^{\tau}e^{\sigma}_{a} \delta e^{\rho \, a} \right) - 3 \, \varepsilon_{\mu \nu \rho \sigma} \, e^{[\sigma}_{a} \delta e^{\rho \, |a|} \nabla_{\tau} \xi^{\tau]}.  
\end{align}
Since we integrate this over a cross-section of $\scri^+$, the first term above is a total derivative; hence it can be neglected. Therefore, 
\begin{align}
\ndelta \mathcal{Q}_\xi &=  \frac{3}{32 \pi G} \int_{\partial \Sigma} \, \varepsilon_{\mu \nu \rho \sigma} \, \left(  g^{\eta [\tau} \xi^{\sigma} \nabla^{\rho]} \delta g_{\eta \tau}-  \, e^{[\sigma}_{a} \delta e^{\rho \, |a|} \nabla_{\tau} \xi^{\tau]} \right)  dx^{\mu} \swedge dx^{\nu}, \notag \\[3mm]
&=\ndelta \mathcal{Q}_{IW}  +\frac{1}{32 \pi G} \int_{\partial \Sigma} \, \varepsilon_{\mu \nu \rho \sigma} \, \Big( - 3 \, e^{[\sigma}_{a} \delta e^{\rho \, |a|} \nabla_{\tau} \xi^{\tau]}\notag \\
&\hspace{52mm} + g^{\tau \sigma} \delta g_{\tau \eta} \nabla^{\eta} \xi^{\rho} + \delta (\log \sqrt{-g}) \nabla^{\rho} \xi^{\sigma} \Big)  dx^{\mu} \swedge dx^{\nu},
\label{integrand2}
\end{align}
where 
\begin{equation}
 \ndelta \mathcal{Q}_{IW} = \frac{1}{32 \pi G} \int_{\partial \Sigma} \, \varepsilon_{\mu \nu \rho \sigma} \, \left(3 \,  g^{\eta [\tau} \xi^{\sigma} \nabla^{\rho]} \delta g_{\eta \tau} - g^{\tau \sigma} \delta g_{\tau \eta} \nabla^{\eta} \xi^{\rho} - \delta (\log \sqrt{-g} )\nabla^{\rho} \xi^{\sigma} \right)  dx^{\mu} \swedge dx^{\nu}
\end{equation}
is the Iyer-Wald charge calculated from the second order formalism \cite{IW} (see also Ref.\ \cite{Compere}).  It is equal to the Barnich-Brandt charge \cite{BB}; see Ref.\ \cite{Compere}. 

Since $\mu \nu = IJ$, this implies that the $\rho \sigma$ indices in the extra terms in equation \eqref{integrand2} must be $[ur]$.  Using equations \eqref{vielbeine}, \eqref{xi} and \eqref{drxiu}, this implies that the extra terms are proportional to 
\begin{equation}
 - 3 \, e^{[u}_{a} \delta e^{r \, |a|} \nabla_{\tau} \xi^{\tau]} + g^{\tau [u} \delta g_{\tau \eta} \nabla^{|\eta|} \xi^{r]} + \delta (\log \sqrt{-g}) \nabla^{[r} \xi^{u]} = e^{- 2 \beta} \delta \beta \left( \nabla_{I} \xi^{I} - C^{I} \nabla_{I} \xi^{u} \right),
\end{equation}
which vanishes by identity \eqref{divxiI}.

Therefore, from equation \eqref{integrand2}
\begin{equation}
 \ndelta \mathcal{Q}_\xi = \ndelta \mathcal{Q}_{IW}.
\end{equation}
In summary, a first order analysis of the Palatini action reproduces the Iyer-Wald expression, which is also equal to the Barnich-Brandt expression, giving rise to the standard leading order BMS charges \cite{BarTro}, as well as the subleading BMS charges \cite{fakenews}. 

\subsubsection{Dual BMS charges}

Now, we turn to the dual BMS charges, which arise from the Holst term in the action,
\begin{equation}
\ndelta \td{\mathcal{Q}}_\xi = \frac{1}{8 \pi G} \int_{\partial \Sigma} \, \iota_{\xi}e_a \, \delta \omega^{ab} \swedge e_b.
\end{equation}
As before, using equation \eqref{omega}, it is fairly simple to show that
\begin{equation}
 \ndelta \td{\mathcal{Q}}_\xi = \frac{1}{8 \pi G} \int_{\partial \Sigma} \,  \xi^{\tau} \nabla_{J}\left( e_{\tau \, a} \delta e_{I}^a \right)\ dx^I \swedge dx^J.
 \label{integrandt}
\end{equation}
Expanding the integrand and using the antisymmetrisation in $IJ$
\begin{align}
 \xi^{\tau} \nabla_{J}\left( e_{\tau \, a} \delta e_{I}^a \right) &= \xi^{\tau} \partial_{J}\left( e_{\tau \, a} \delta e_{I}^a \right) - \xi^{\tau} \Gamma^\rho_{J \tau} \left( e_{\rho \, a} \delta e_{I}^a \right) \notag\\
 &= - (\partial_{J} \xi^{\tau} + \Gamma^\tau_{J \rho} \xi^{\rho}) \left( e_{\tau \, a} \delta e_{I}^a \right) \notag \\
 &= - \nabla_{J} \xi^\tau \left( e_{\tau \, a} \delta e_{I}^a \right),
\end{align}
where in the second equality, we have integrated by parts and ignored the total derivative term, which is trivial.  Therefore,
\begin{equation} \label{dualq}
 \ndelta \td{\mathcal{Q}}_\xi = \frac{1}{8 \pi G} \int_{\partial \Sigma} \, \, e_{\tau \, a} \, \delta e_{J}^a \, \nabla_{I} \, \xi^{\tau} \ dx^I \swedge dx^J.
\end{equation}
Consider
\begin{align}
 e_{\tau \, a} \, \delta e_{J}^a \, \nabla_{I} \, \xi^{\tau} & \ dx^I \swedge dx^J = e_{K\, i} \delta e_{J}^i\, \left(\nabla_I \xi^K - C^K \nabla_I \xi^u\right)  \ dx^I \swedge dx^J \notag \\[3mm]
 &= \left(\frac12 \delta g_{KJ} + e_{[K\, |i|} \delta e_{J]}^i\right) \left(\nabla_I \xi^K - C^K \nabla_I \xi^u \right)  \ dx^I \swedge dx^J \notag \\[3mm]
 &= \frac12 \left\{ \delta g_{KJ} \left(\nabla_I \xi^K - C^K \nabla_I \xi^u \right) +  e_{I\, i} \delta e_{J}^i \left(\nabla_K \xi^K - C^K \nabla_K \xi^u \right) \right\}  dx^I \swedge dx^J \notag \\[3mm]
 &= \frac12 \delta g_{KJ} \left(\nabla_I \xi^K - g^{rK} \nabla_I \xi_r \right) dx^I \swedge dx^J,
\end{align}
where in the first equality we have used \eqref{vielbeine}, in the third equality we have used a Schouten identity and in the fourth equality we have used identity \eqref{divxiI}, as well as the form of the inverse metric, which gives that $g^{rK} = g^{ur} C^K.$  Inserting the above equality into equation \eqref{dualq} gives\footnote{In this subsection, we are lowering and raising all indices with $g_{\mu \nu}$ and its inverse, including $IJ$ indices. Therefore, here $\xi_I = g_{I \mu} \xi^\mu$.}
\begin{align}
 \ndelta \td{\mathcal{Q}}_\xi &= \frac{1}{16 \pi G} \int_{\partial \Sigma} \, \, \delta g_{KJ} \left(\nabla_I \xi^K - g^{rK} \nabla_I \xi_r \right) \ dx^I \swedge dx^J. \notag \\[3mm]
 &= \frac{1}{32 \pi G} \int_{\partial \Sigma} \, \delta g_{JK} \left( \nabla_{I} \xi^{K} + \nabla^{K} \xi_{I} \right)  \ dx^I \swedge dx^J
 + \frac{1}{16 \pi G} \int_{\partial \Sigma} \, d_{IJ} \ dx^I \swedge dx^J, \label{dslashQtp}
\end{align}
where the first expression is the dual charge proposed in Ref.\ \cite{dualex}\footnote{The dual charge is defined in equation (3.1), (3.2) of Ref.\ \cite{dualex} and is equal to the above expression up to a trivial total derivative.} and the difference between the two charges is proportional to the integral of
\begin{align}
 d_{IJ} &= \delta g_{K[J} \left(\nabla_{I]} \xi^K - g^{rK} \nabla_{I]} \xi_r \right) - \frac12 \delta g_{K [ J} \left( \nabla_{I]} \xi^{K} + \nabla^{K} \xi_{I]} \right) \notag \\[3mm]
 &= \frac12 g^{K \tau} \delta g_{K [ J} \left( \nabla_{I]} \xi_{\tau} - \nabla_{|\tau|} \xi_{I]} \right) - g^{rK} \delta g_{K[J} \nabla_{I]} \xi_r \notag \\[3mm]
 &= \frac32 g^{K \tau} \delta g_{K [ J} \nabla_{I} \xi_{\tau]} - g^{rK} \delta g_{K[J} \nabla_{I]} \xi_r \notag \\[3mm]
 &= \frac12 g^{rK} \delta g_{K [J} \left( \nabla_{I]} \xi_{r} - \nabla_{|r|} \xi_{I]} \right) - g^{rK} \delta g_{K[J} \nabla_{I]} \xi_r \notag \\[3mm]
 &=0,
\end{align}
where in the third equality we have used equation \eqref{con:det} and the fact that $\delta \, (\textup{det} \hat{E}^i_I) = 0$ and in the final equality we have again used identity \eqref{symrI}.  

In summary,
\begin{equation}
 \ndelta \td{\mathcal{Q}}_\xi = \frac{1}{32 \pi G} \int_{\partial \Sigma} \, \delta g_{JK} \left( \nabla_{I} \xi^{K} + \nabla^{K} \xi_{I} \right)  \ dx^I \swedge dx^J,
\end{equation}
reproducing the dual BMS charges \cite{dual0} as well as the subleading dual BMS charges \cite{dualex}.  

\subsection{Lorentz transformations} \label{subsec:lorentz}

In addition to diffeomorphisms, there exist another set of non-trivial transformations in the first order formalism; that of Lorentz transformations parametrised by $\Lambda$.  The asymptotic symmetry analysis implies that the set of Lorentz transformations that preserve the boundary conditions, and can thus be viewed as improper gauge transformations, are those given in \eqref{lorentz}.  In this section, we consider what the asymptotic charges associated with these transformations are.  

Applying the general discussion in section \ref{sec:covrev} to Lorentz transformations, we find that the asymptotic charge is defined as
\begin{equation} \label{Lor:Q}
 \ndelta H_\Lambda = \int_\Sigma \omega(\phi, \delta \phi, \delta_\Lambda \phi),
\end{equation}
where
\begin{equation}
  \omega(\phi, \delta \phi, \delta_\Lambda \phi) = \delta \theta(\phi, \delta_\Lambda \phi) - \delta_\Lambda \theta(\phi, \delta \phi),
\end{equation}
where $\theta(\phi,\delta \phi)$ is given in equation \eqref{EHH:sympot}.  The Lorentz transformation acts on the fields as
\begin{equation} \label{lor:trans}
 \delta_\Lambda e^a = \Lambda^a{}_b\, e^b, \qquad \delta_\Lambda \omega^{ab} = -d \Lambda^{ab} + [\Lambda, \omega]^{ab}.
\end{equation}
Consider 
\begin{align}
 \delta_\Lambda \theta(\phi, \delta \phi) &= \frac{1}{16 \pi G} P_{abcd} \left\{ 2 \Lambda^a{}_e\, e^e \swedge e^b \swedge \omega^{cd} + e^a \swedge e^b \swedge \delta \left(-d \Lambda^{cd} + [\Lambda, \omega]^{cd}\right) \right\} \notag \\
 &= \frac{1}{8 \pi G} P_{abcd} \left\{\Lambda^a{}_e\, e^e \swedge e^b \swedge \omega^{cd} +\Lambda^c{}_e\, e^a \swedge e^b \swedge \delta \omega^{ed} \right\} \notag \\
 &=0, \label{lor:lth}
\end{align}
where in the first equality we have used equations \eqref{EHH:sympot} and \eqref{lor:trans}, in the second equality we have used that $\delta \Lambda = 0$ and the third equality results from a Schouten identity.  Furthermore, it is simple to show that
\begin{equation} \label{lor:thl}
 \theta(\phi, \delta_{\Lambda} \phi) = d Q_{\Lambda}(\phi),
\end{equation}
where
\begin{equation} \label{lor:N}
  Q_{\Lambda}(\phi) = \frac{1}{16 \pi G} P_{abcd} \, \Lambda^{ab} \,e^c\swedge e^d.
\end{equation}
Therefore, using equations \eqref{lor:lth} and \eqref{lor:thl}, equation \eqref{Lor:Q} simplifies to
\begin{equation}
 \ndelta H_\Lambda = \int_{\partial \Sigma} \delta Q_{\Lambda}(\phi).
\end{equation}
The components of $Q_{\Lambda}(\phi)$ that the integral above projects to are its $IJ$ components.  From equation \eqref{lor:N},
\begin{equation}
 {Q_\Lambda}_{IJ} = - \frac{r^2}{16 \pi G} P_{abij}\, \Lambda^{ab}\, \varepsilon^{ij}\, \varepsilon_{IJ},
\end{equation}
where $\varepsilon_{IJ}$ is the volume form on the round 2-sphere (see appendix \ref{app:twist}).  In order to obtain the above expression, importantly, we have used the determinant condition \eqref{con:det}.  Clearly, the variation of the right hand side of the above expression is zero, which implies that
\begin{equation}
 \ndelta H_\Lambda = 0,
\end{equation}
i.e.\ asymptotic Lorentz transformations lead to trivial asymptotic charges.  One way to understand this result is that Lorentz transformations correspond to degenerate directions in phase space.  Using some local coordinates $A, B, \ldots$ on phase space, recall that degenerate directions correspond precisely to those transformations $X$ such that
\begin{equation} 
 \omega_{AB} X^B = 0.
\end{equation}
Thus, what we thought were large Lorentz gauge transformations turned out to be proper; consequently leading to a trivial charge.

\section{Identifying the integrable charge}
\label{sec:int}

We explained towards the end of section \ref{sec:covrev} how diffeomorphism charges are, in general, not integrable. As illustrated in equation \eqref{Q:intnint}, $\ndelta H_{\xi}$ can be split into two terms: the variation of an integrable charge $\mathcal{H}_\xi$ and a non-integrable term $\mathcal{N}_\xi$.  The physics behind the existence of such a non-integrable terms is clear; it is related to the existence of flux at null infinity removing charge from the spacetime.  As such equation \eqref{Q:intnint} can be viewed as a generalised continuity equation in the following sense (see also ref.\ \cite{dualex}):  Given the properties of the asymptotic charge, on-shell
\begin{equation}
 \ndelta H_\xi (\phi,\delta_\xi \phi) = 0.
\end{equation}
Therefore, in this case, equation \eqref{Q:intnint} implies that the change in the integrable charge is balanced by the change in flux; this is a continuity equation.  However, an important issue that arises when defining the splitting in order to derive an integrable charge is how to physically fix the ambiguity \eqref{Q:amb}.  This issue is the object of attention of Wald-Zoupas \cite{WZ} and what they find is that for standard BMS charges at leading order, the prescription that should be followed is to pull-back the presymplectic 2-form to infinity, read off the associated potential, what they call $\Theta$ and subtract this from the $\theta$ term in the definition of the charge; see equations \eqref{bigTh} and \eqref{intQ}.  This makes sense, because the non-integrability comes from the existence of the $\theta$ term in the expression for the charge \eqref{diffQ} and the pull-back of the presymplectic 2-form to infinity parametrises the flux at infinity.  Therefore, it is natural to remove the contribution of potential $\Theta$ associated with the pull-back of the presymplectic 2-form from the expression involving $\theta$ in order to determine the integrable charge.

In this section, we show that the Wald-Zoupas prescription also works in the first order formalism to leading order and that it determines in particular the leading order integrable dual charge.  Following Ref.\ \cite{WZ}, we begin by considering the pull-back of the presymplectic 2-form to a constant $r$ surface, i.e.\ we consider its $uIJ$ component
\begin{equation} \label{pullback}
 \bar{\omega}(\phi, \delta_1 \phi, \delta_2 \phi) = \frac{1}{16 \pi G}\, P_{abcd} \, \delta_1 \left(e^a\swedge e^b \swedge \delta_2 \omega^{cd}\right) - (1 \leftrightarrow 2).
\end{equation} 
Consider the Hodge dual of the presymplectic form
\begin{equation} \label{Hdual}
 (\star \omega)^\mu = \frac{1}{6}\, \varepsilon^{\mu \nu \rho \sigma} \omega_{\nu \rho \sigma}, \qquad \omega_{\mu \nu \rho} = \varepsilon_{\mu \nu \rho \sigma} (\star \omega)^\sigma.
\end{equation}
The pull-back of the presymplectic 2-form to a constant $r$ surface implies that we consider 
\begin{align} 
 (\star \omega)^r &= \frac{1}{8 \pi G}\, \varepsilon^{r \nu \rho \sigma}\, P_{abcd} \, \delta_1 e^a_{[\nu}\, e^b_{\rho}\, \delta_2 \omega^{cd}_{\sigma]} - (1 \leftrightarrow 2) \notag \\[3mm]
 &=  \frac{3}{8 \pi G}\, \left( e_a^{[r} e_c^{\nu} e_d^{\sigma]}\, \delta_1 e^a_{\nu}\, \delta_2 \omega^{cd}_{\sigma} 
 + i \lambda\, r^{-2} \, e^{-2\beta} \, \varepsilon^{IJ}\, \delta_1 e^a_{[u}\, e^b_{I}\, \delta_2 \omega_{J]ab} \right) - (1 \leftrightarrow 2) \notag \\[3mm]
 &=  \frac{1}{8 \pi G}\, \left( \left[e_c^{\nu} (e_a^{r}  e_d^{\sigma} - e_a^{\sigma}  e_d^{r}) -  e_c^{r} e_d^{\sigma} e^\nu_a \right]  \delta_1 e^a_{\nu}\, \delta_2 \omega^{cd}_{\sigma} \right. \notag \\[2mm]
 & \hspace{12mm} \left. + i \lambda\, r^{-2} \, e^{-2\beta} \, \varepsilon^{IJ}\, \left[ \delta_1 e^a_{u}\, e^i_{I}\, \delta_2 \omega_{Jai} - \delta_1 e^i_{I}\, e^b_{u}\, \delta_2 \omega_{Jib} 
 + \delta_1 e^i_{I}\, e^j_{J}\, \delta_2 \omega_{uij} \right] \right) - (1 \leftrightarrow 2) \notag \\[3mm]
 &=  \frac{1}{8 \pi G}\, \left( - \left( \delta_1 e_c^{r}\,  e_d^{\sigma} - \delta_1 e_c^{\sigma}\,  e_d^{r} \right) \, \delta_2 \omega^{cd}_{\sigma} -  2 \delta_1 \beta \,  e_c^{r} e_d^{\sigma}\, \delta_2 \omega^{cd}_{\sigma} \right. \notag \\[2mm]
 & \hspace{20mm} \left. + i \lambda\, r^{-2} \, e^{-2\beta} \, \varepsilon^{IJ}\, \left[ \delta_1 (e^a_{u}\, e^i_{I}) \, \delta_2 \omega_{Jai} + \delta_1 e^i_{I}\, e^j_{J}\, \delta_2 \omega_{uij} \right] \right) - (1 \leftrightarrow 2),
 \label{Hdualr}
\end{align}
where in the first equality we have substituted equations \eqref{Hdual} and \eqref{pullback}; in the second equality we have used equation \eqref{P} and that $\varepsilon^{urIJ} = - r^{-2} e^{-2\beta} \varepsilon^{IJ}$; in the third equality we have used the definition of the vierbein \eqref{vielbeine} and in the fourth equality we have used the fact that 
$\textup{det} (e^a_\mu) = r^2 e^{2\beta} \textup{det} (\hat{E}_I^i)$.  From the expressions for the spin connection \eqref{spincon}, it is fairly simple to see that
\begin{align*}
 \delta \omega_{01} &= O(r^{-2}) du \ + O(r^{-2}) dr \ + O(r^{-1}) dx^I, \quad \delta \omega_{0i} = O(r^{-2}) du \ + O(r^{-1}) dx^I, \notag \\[2mm]
 \delta \omega_{1i} &= O(r^{-2}) du \ + O(r^{-2}) dr \ + O(r^{0}) dx^I, \quad \delta \omega_{ij} = O(r^{-1}) du \ + O(r^{-1}) dr \ + O(r^{-1}) dx^I.
\end{align*}
Using the above expressions and the form of the vierbein \eqref{vielbeine}, \eqref{falloffs}, equation \eqref{Hdualr} becomes
\begin{align}
 (\star \omega)^r &= \frac{1}{8 \pi G\, r}\, \left( \delta_1 E^{iI} \, \delta_2 \omega_{I1i} + i \lambda \, \varepsilon^{IJ}\,\delta_1 E^i_{I} \left[ \, \delta_2 \omega_{J1i} + r E^j_{J}\, \delta_2 \omega_{uij} \right] + o(r^{-1}) \right) - (1 \leftrightarrow 2) \notag \\[3mm]
 &= - \frac{1}{8 \pi G}\, \left(\delta_1 E^{iI} \, \delta_2 \left[ E^J_{(i} \partial_{|u|} E_{j)J} E^j_I \right]  + i \lambda\, \varepsilon^{IJ}\,\delta_1 E^i_{I}\, \delta_2 \partial_u E_{iJ} + o(r^{-2}) \right) - (1 \leftrightarrow 2) \notag \\[3mm]
&= - \frac{1}{8 \pi G}\, \left(\frac14 \delta_1 h^{IJ} \, \delta_2 \partial_u h_{IJ}  + i \lambda\, \varepsilon^{IJ}\,\delta_1 E^i_{I}\, \delta_2 \partial_u E_{iJ} + o(r^{-2}) \right) - (1 \leftrightarrow 2).
 \end{align}
Now, from equation \eqref{Hdual},
\begin{align}
 \omega_{uIJ} &= \varepsilon_{uIJr} (\star \omega)^r \notag \\[3mm]
 &= \frac{r^2 \varepsilon_{IJ}}{8 \pi G}\, \delta_1 \left(\frac14 \delta_2 h^{KL} \, \partial_u h_{KL}  + i \lambda\, \varepsilon^{KL}\,\delta_2 E^i_{K}\, \partial_u E_{iL} + o(r^{-2}) \right) - (1 \leftrightarrow 2).
\end{align}
Using the expansion for $E^i_I$ in equation \eqref{falloffs} and the fact that
\begin{equation}
 h_{IJ} = \gamma_{IJ} + \frac{C_{IJ}}{r} + o(r^{-1}), \qquad h^{IJ} = \gamma^{IJ} - \frac{C^{IJ}}{r} + o(r^{-1}),
\end{equation}
\begin{align}
 \omega_{uIJ} &= - \frac{\varepsilon_{IJ}}{32 \pi G}\, \delta_1 \left(\delta_2 C^{KL} \, \partial_u C_{KL}  + i \lambda\, \delta_2 \widetilde{C}^{KL} \, \partial_u C_{KL} + o(r^{0}) \right) - (1 \leftrightarrow 2),
\end{align}
where the twist of tensors on the round 2-sphere are defined in appendix \ref{app:twist}.
Using equation \eqref{bigTh}, we conclude that at leading order
\begin{equation}
 \Theta_{uIJ}^{(0)} = - \frac{\varepsilon_{IJ}}{32 \pi G}\, \left(\delta C^{KL} \, \partial_u C_{KL}  + i \lambda\, \delta \widetilde{C}^{KL} \, \partial_u C_{KL}  \right).
\end{equation}
  Therefore, the leading order non-integrable part of the variation of the asymptotic charges, as defined in equation \eqref{nonint}, is equal to
\begin{equation}
 \mathcal{N}_\xi^{(0)} = \frac{1}{32 \pi G} \int_{\partial \Sigma} d\Omega\;  \xi^u \left(\delta C^{KL} \, \partial_u C_{KL}  + i \lambda\, \delta \widetilde{C}^{KL} \, \partial_u C_{KL}  \right),
\end{equation}
where $d\Omega$ is the volume form on the unit round 2-sphere.  This matches that expected from previous studies \cite{WZ, BarTro, dual0, dualex}.

What remains is to prescribe a similar procedure for finding the subleading integrable charges.  Note that whereas null infinity may be viewed as a $r=constant$ surface, subleading charges will live away from null infinity and as such will live on $v=constant$ null surfaces, where $v$ is the ingoing Eddington-Finkelstein-like timelike coordinate.  However, pulling the presymplectic 2-form to $v=constant$ surfaces does not lead to a sensible answer.  While, it is simple to distinguish the integrable charge at subleading orders on a case by case basis \cite{fakenews, dualex}, it is clear that a Wald-Zoupas-like prescription that determines the subleading integrable charge in a general, geometric, way by pulling the presymplectic 2-form to some surface is more challenging.  We hope to deal with this interesting problem in future work. 

\section{Charge algebra for leading order dual charges} \label{sec:algebra}

In this section, we derive the charge algebra associated with leading order dual charges and show that they satisfy the same algebra as the standard leading BMS charges, albeit with a slightly different, but analogous, field dependent central extension.  The leading order dual BMS charge corresponding to the full BMS algebra is \cite{dualex}\footnote{Note that there is a minor typographical error in equation (4.6) of Ref.\ \cite{dualex}.}
\begin{equation}
 \ndelta \widetilde{\mathcal{Q}}_{0\, \xi} = \delta \widetilde{\mathcal{Q}}^{(int)}_{0\, \xi} + \widetilde{\mathcal{N}}_{0 \, \xi}[\delta \phi]
\end{equation}
with
\begin{gather}
\widetilde{\mathcal{Q}}^{(int)}_{0\, \xi} =\frac{1}{16 \pi G} \int_{\partial \Sigma}\, d\Omega \, \Bigg( - f D_I D_J \widetilde{C}^{IJ}  + \frac{1}{4} Y^K \widetilde{C}^{IJ} D_K C_{IJ} - \frac{1}{4} \widetilde{Y}^I D_I  C^2 \Bigg), \label{dual0:int} \\[3mm]
\widetilde{\mathcal{N}}_{0\, \xi}[\delta \phi] =\frac{1}{32 \pi G} \int_{\partial \Sigma}\, d\Omega \ f\, \partial_u C_{IJ}\, \delta \widetilde{C}^{IJ}. \label{dual0:nint}
\end{gather}

Following Ref.\ \cite{BarTro}, we define the bracket of the charges to be\footnote{Note that the relative minus sign difference with Ref.\ \cite{BarTro} in the definition of the bracket is due to the difference in defining the action of the BMS generators on the metric components.  This difference can be traced back to whether one views BMS transformations as acting actively or passively on the fields.}
\begin{equation}
 \{ \widetilde{\mathcal{Q}}^{(int)}_{0\, \xi_1}, \widetilde{\mathcal{Q}}^{(int)}_{0\, \xi_2} \} = \delta_{\xi_2} \widetilde{\mathcal{Q}}^{(int)}_{0\, \xi_1} + \widetilde{\mathcal{N}}_{0 \, \xi_2}[\delta_{\xi_1} \phi].
\end{equation}
Inspecting equations \eqref{dual0:int} and \eqref{dual0:nint}, clearly the only relevant field transformations are those acting on $C_{IJ}$, which transforms in the following way\footnote{See, for example, equation (2.18) of Ref.\ \cite{BarTro}.}
\begin{equation} \label{var:C}
 \delta C_{IJ} = f \partial_u C_{IJ} + \Box f\, \gamma_{IJ} - 2 D_{(I} D_{J)} f + Y^K D_K C_{IJ} + 2 C_{K(I} D_{J)} Y^K - \half D_K Y^K\, C_{IJ}.
\end{equation}
Consequently, it is simple to show that
\begin{equation} \label{var:Ct}
 \delta \widetilde{C}^{IJ} = f \partial_u \widetilde{C}^{IJ} + 2 \varepsilon^{K(I} D_{K} D^{J)} f + Y^K D_K \widetilde{C}^{IJ} + 2 \widetilde{C}_K{}^{(I} D^{J)} Y^K - \half D_K Y^K\, \widetilde{C}^{IJ}
\end{equation}
and
\begin{equation} \label{var:C2}
 \delta C^2 = f \partial_u C^2 -4 C^{IJ} D_{I} D_{J} f + D_K \left( C^2 Y^K \right).
\end{equation}
Using the above expressions and making extensive use of the fact that $Y^I$ is a conformal Killing vector on the round 2-sphere, see equation \eqref{Y:CKV}, as well as Schouten identities described in appendix B of \cite{fakenews}, one can show that\footnote{See appendix \ref{app:algebra} for a detailed derivation of this result.} 
\begin{equation} \label{Alg:1}
  \{ \widetilde{\mathcal{Q}}^{(int)}_{0\, \xi_1}, \widetilde{\mathcal{Q}}^{(int)}_{0\, \xi_2} \} = \widetilde{\mathcal{Q}}^{(int)}_{0\, [\xi_1, \xi_2]} + \widetilde{K}_{\xi_1,\xi_2},
\end{equation}
where the commutation of two BMS generators $[\xi_1, \xi_2]$ corresponds to a third BMS generator $\xi_3$ with \cite{BarTro}
\begin{equation}
 f_3 = Y_1^I D_I f_2 - \half f_2\, D_K Y_1^K - Y_2^I D_I f_1 + \half f_1\, D_K Y_2^K, \quad \ Y_3^I = Y_1^K D_K Y_2^I - Y_2^K D_K Y_1^I.
\end{equation}
The field dependent central extension
\begin{equation} \label{Alg:2}
 \widetilde{K}_{\xi_1,\xi_2} = \frac{1}{32 \pi G} \int_{\partial \Sigma}\, d\Omega \ \ \widetilde{C}^{IJ} \Big( f_1\, D_I D_J \, D_K Y_2^K - f_2\, D_I D_J \, D_K Y_1^K \Big).
\end{equation}
Compare this with the field dependent central extension corresponding to the leading order BMS charges \cite{BarTro}
\begin{equation}
 K_{\xi_1,\xi_2} = \frac{1}{32 \pi G} \int_{\partial \Sigma}\, d\Omega \ \ C^{IJ} \Big( f_1\, D_I D_J \, D_K Y_2^K - f_2\, D_I D_J \, D_K Y_1^K \Big).
\end{equation}

\section{Fermions} \label{sec:ferm}

In section \ref{sec:charges}, we computed the asymptotic charges corresponding to asymptotically flat solutions of the Palatini-Holst theory \eqref{EHH}, i.e.\ Einstein gravity in the first order formalism with an extra term, called the Holst term, that does not contribute to the equations of motion and hence its existence at the level of the action cannot be ruled out.  A lot of what we found for this theory relied heavily on the fact that the torsion vanished as a result of the equation of motion for the spin connection.  The fact that the Holst term does not contribute to the equations of motion, for example, is itself a consequence of the fact that the torsion vanishes.  

In this section, we assess the extent to which similar results as in section \ref{sec:charges} may be obtained in the case where there exists non-trivial torsion, which is the subject of Einstein-Cartan theory \cite{cartan, Kibble:1961ba, hehl}.  A simple situation in which torsion arises is in the presence of fermions.  Therefore, in this section, we consider asymptotic charges in a setting in which one has gravity as well as fermions.  We will find that asymptotic charges, including dual charges, can still be defined, following some minor, yet important, modifications. The results of this section were already reported in \cite{letter}.

As remarked above, in the presence of torsion, the Holst term is no longer trivial in terms of its contribution to the equation of motion (the Einstein equation).  Consequently, it must be modified.  The analogous term is the Nieh-Yan term \cite{NiehYan}
\begin{equation}
  S_{NY} = \frac{i \lambda}{16 \pi G}  \int_{\mathcal M} \left( \mathcal{R}_{ab}(\omega) \swedge e^{a} \swedge e^{b} - T^a \swedge T_a \right).
 \label{NY}
\end{equation}
Using the fact that in the presence of torsion, Cartan's first structure equation \eqref{Cartan1} becomes
\begin{equation}
 d e^a + \omega^a{}_{b} \swedge e^b=T^a,
 \label{T:Cartan1}
\end{equation}
and the algebraic Bianchi identity becomes
\begin{equation}
 d T^a + \omega^a{}_{b} \swedge T^b= \mathcal{R}^a{}_{b} \swedge e^b,
 \label{T:aBianchi}
\end{equation}
it is fairly simple to show that 
\begin{equation}
 \mathcal{R}_{ab}(\omega) \swedge e^{c} \swedge e^{d} - T^a \swedge T_a = - d \left( e^a \swedge T_a \right).
\end{equation}
Therefore, the Nieh-Yan term can be written as an exact term.  In this form it is clearer to see that it vanishes in the absence of torsion, as a result of the algebraic Bianchi identity.  In order, to maintain the connection with section \ref{sec:charges}, we want to view the Nieh-Yan term as a correction to the Holst term in the presence of torsion.  Accordingly, we use the form of the Nieh-Yan term given in equation \eqref{NY}, rather than its simpler exact form.  Adding this term \eqref{NY} to the Palatini-Dirac action gives
\begin{equation}
 S_{PNYD} = \frac{1}{16 \pi G}  \int_\mathcal{M} \left( P_{abcd} \, \mathcal{R}^{ab}(\omega) \swedge e^{c} \swedge e^{d} - i \lambda T^a \swedge T_a  \right) + \frac{1}{2} \int_\mathcal{M} \varepsilon\, \overline{\psi} \overleftrightarrow{\slashed{\nabla}} \psi,
 \label{EHNYD}
\end{equation}
where $P_{abcd}$ is defined in equation \eqref{P}, $\varepsilon$ denotes the volume form, 
\begin{equation}
 \overline{\psi} = i \psi^\dagger \gamma^0, \qquad \{\gamma^a \gamma^b\} = 2 \eta^{ab}
\end{equation}
and the operator
\begin{equation}
 \overleftrightarrow{\nabla} = \overrightarrow{\nabla} - \overleftarrow{\nabla}, \qquad \slashed\nabla \equiv e_a^\mu \gamma^a \nabla_\mu
\end{equation}
with the covariant derivative acting on spinors as
\begin{equation}
 \nabla_{\mu} \psi = \partial_{\mu} \psi + \frac14 \omega_{\mu}{}^{ab} \gamma_{ab} \psi. 
\end{equation}
Varying action \eqref{EHNYD} with respect to $\psi$ gives the Dirac equation
\begin{equation}
 \slashed{\nabla} \psi =0,
 \label{dirac}
\end{equation}
while varying with respect to $\omega$, we obtain
\begin{equation}
 \frac{1}{8 \pi G} \left( P_{abcd} [ d e^c + \omega^{c}{}_{e} \swedge e^{e}] \swedge e^d   - i \,\lambda \,  T_{[a} \swedge e_{b]} \right) + \frac{1}{24} \, \varepsilon_{cdef} \, \overline{\psi} \gamma^{cde} \psi \  e_a \swedge e_b \swedge e^f = 0,
\end{equation}
which using Cartan's first structure equation \eqref{T:Cartan1} reduces to 
\begin{equation}
 \frac{1}{16 \pi G} \varepsilon_{abcd} T^c \swedge e^d   + \frac{1}{24} \, \varepsilon_{cdef} \, \overline{\psi} \gamma^{cde} \psi \  e_a \swedge e_b \swedge e^f  = 0.
\end{equation}
This determines the torsion in terms of the Dirac fields
\begin{equation} \label{torsion}
 T^a = -2\pi G  \ \overline{\psi} \gamma^{abc} \psi \  e_b \swedge e_c.
\end{equation}
The Einstein equation, obtained by varying the vierbein, is\footnote{Note that we have used the Dirac equation \eqref{dirac} to simplify the resulting expression.}
\begin{equation}
 \frac{1}{8 \pi G} \varepsilon_{abcd} \mathcal{R}^{ab} \swedge e^c   +  e^{\nu}_{d} \left( \overline{\psi} \, \iota_{\gamma} \varepsilon  \, \nabla_{\nu} \psi - \overline{\psi} \overleftarrow{\nabla}_{\nu}\, \iota_{\gamma} \varepsilon  \,  \psi \right) 
 = 0,
 \label{einstein}
\end{equation}
where
\begin{equation}
 \iota_{\gamma} \varepsilon = \frac{1}{6} \gamma^{a} \varepsilon_{abcd} \, e^b \swedge e^c \swedge e^d.
\end{equation}
Equivalently, 
\begin{equation}
 G^\mu{}_\nu + 4 \pi G  \, e^\mu_a \, \left( \overline{\psi} \gamma^a \nabla_{\nu} \psi - \overline{\psi} \overleftarrow{\nabla}_{\nu} \gamma^a \psi \right) = 0,
\end{equation}
where $G^\mu{}_\nu = R^\mu{}_\nu - \frac12 R\, \delta^\mu_\nu$ is the Einstein tensor.

The presymplectic potential corresponding to theory \eqref{EHNYD} is 
\begin{equation}
 \theta(\phi, \delta \phi) = \frac{1}{16 \pi G} \left( P_{abcd} \, \delta \omega^{ab} \swedge e^c \swedge e^d - 2  i  \lambda \, \delta e^a \swedge T_{a} \right) +  \frac{1}{2} \left( \overline{\psi} \, \iota_{\gamma} \varepsilon \, \delta \psi - \delta\overline{\psi} \, \iota_{\gamma} \varepsilon  \,  \psi \right),
 \label{presympot}
 \end{equation}
while, the Noether charge, as defined by equation \eqref{current} is 
 \begin{equation}
  Q= \frac{1}{16 \pi G} \left( P_{abcd} \,   \iota_{\xi}\omega^{ab} \,  e^c \swedge e^d - 2  i  \lambda \, \iota_{\xi} e^a \, T_{a} \right).
 \end{equation}
We can verify that the Noether charge as defined above does indeed satisfy equation \eqref{current} by taking the exterior derivative of the expression above, using Cartan's magic formula \eqref{magic} and Schouten identities to find that
 \begin{align}
  dQ= \frac{1}{16 \pi G} & \left( P_{abcd} \,   \mathcal{L}_{\xi}\omega^{ab} \swedge  e^c \swedge e^d - 2  i  \lambda \ \mathcal{L}_{\xi} e^a \swedge T_{a} \right)  - \iota_{\xi} L  \notag \\[5pt]
  & - \frac{1}{32 \pi G} \varepsilon_{abcd} \left( e^a \swedge e^b \swedge \iota_{\xi}\mathcal{R}^{cd} - 2\, T^a \swedge e^b\ \iota_{\xi} \omega^{cd} \right). \label{dQ}
  \end{align}
Consider the terms on the second line of the right hand side above
\begin{align*}
- \frac{1}{32 \pi G} & \varepsilon_{abcd} \left( e^a \swedge e^b \swedge \iota_{\xi}\mathcal{R}^{cd} - 2\, T^a \swedge e^b\ \iota_{\xi} \omega^{cd} \right) \\[5pt]
 = \, & \frac{1}{32 \pi G} \varepsilon_{abcd} \left( e^a \swedge \iota_{\xi} \left[ e^b \swedge \mathcal{R}^{cd} \right] - e^a \swedge \mathcal{R}^{cd} \, \iota_{\xi} e^b + 2\, T^a \swedge e^b\ \iota_{\xi} \omega^{cd} \right)  \\[5pt]
 = \, & \frac{1}{12} \varepsilon_{abcd} \left( \xi^{\mu} \left[ \overline{\psi} \, \gamma^{a}  \, \nabla_{\mu} \psi - \overline{\psi} \overleftarrow{\nabla}_{\mu}\, \gamma^a   \,  \psi \right] - \frac{1}{2}\, \iota_{\xi} \omega^{ef} \, \overline{\psi} \gamma^{a}{}_{ef}  \psi \right) e^b \swedge e^c \swedge e^d  \\[5pt]
 = \, & \frac{1}{2} \left( \overline{\psi} \, \iota_{\gamma} \varepsilon \, \mathcal{L}_{\xi} \psi - \mathcal{L}_{\xi}\overline{\psi} \, \iota_{\gamma} \varepsilon  \,  \psi \right),
\end{align*}
where 
\begin{equation}
 \mathcal{L}_{\xi} \psi = \xi^{\mu} \partial_{\mu} \psi.
\end{equation}
In the penultimate equality we have used the Einstein equation \eqref{einstein} and the expression for the torsion given in equation \eqref{torsion}.  Therefore, from equation \eqref{dQ} and the definition of the presymplectic potential \eqref{presympot}, we establish 
\begin{equation}
   dQ= \theta(\phi, \mathcal{L}_{\xi} \phi) - \iota_{\xi} L.
\end{equation}
The variation of the asymptotic charge is given by equation \eqref{diffQ}, hence we consider on the sphere
\begin{align}
 \delta Q - \iota_\xi \theta = \frac{1}{8 \pi G} P_{abcd} \ \iota_{\xi}e^c\ \delta \omega^{ab} \swedge e^d 
 - \frac{i \lambda}{8 \pi G}& \left( \iota_\xi e^a \ \delta T_a + \delta e^a \swedge \iota_\xi T_a \right) \notag \\[3mm]
 &- \frac{1}{2} \left( \overline{\psi} \, \iota_\xi \iota_{\gamma} \varepsilon \, \delta \psi - \delta \overline{\psi} \, \iota_\xi \iota_{\gamma} \varepsilon  \,  \psi \right),
 \label{T:ndH}
\end{align}
where we have used equation \eqref{vee} to simplify the expression on the right hand side.  Again, using equation \eqref{vee} and ignoring total derivative terms, it is simple to show from the definition of the torsion \eqref{T:Cartan1} that 
\begin{equation}
 \iota_\xi e^a \ \delta T_a + \delta e^a \swedge \iota_\xi T_a = \delta e^a \swedge \mathcal{L}_{\xi} e_a + \iota_{\xi}e^a\ \delta \omega_{ab} \swedge e^b.
\end{equation}
Substituting the above equation into equation \eqref{T:ndH} and using the definition \eqref{P}, on the sphere
\begin{align}
  \delta Q - \iota_\xi \theta = \frac{1}{16 \pi G} \varepsilon_{abcd} \ \iota_{\xi}e^c\ \delta \omega^{ab} \swedge e^d - \frac{1}{2} \left( \overline{\psi} \, \iota_\xi \iota_{\gamma} \varepsilon \, \delta \psi - \delta \overline{\psi} \, \iota_\xi \iota_{\gamma} \varepsilon  \,  \psi \right) - \frac{i \lambda}{8 \pi G} \delta e^a \swedge \mathcal{L}_{\xi} e_a.
\end{align}
In summary, the presence of torsion does not impede the definition of dual gravitational charges and, in particular, for the Einstein-Dirac theory, we have that 
\begin{equation}
 \ndelta H^{(T)}_{\xi} = \ndelta \mathcal{Q}^{(T)}_\xi + i\, \lambda \, \ndelta \td{\mathcal{Q}}^{(T)}_\xi,
\end{equation}
where
\begin{gather}
\ndelta \mathcal{Q}^{(T)}_\xi = \int_{\partial \Sigma} \left\{ \frac{1}{16 \pi G} \varepsilon_{abcd} \ \iota_{\xi}e^c\ \delta \omega^{ab} \swedge e^d - \frac{1}{2} \left( \overline{\psi} \, \iota_\xi \iota_{\gamma} \varepsilon \, \delta \psi - \delta \overline{\psi} \, \iota_\xi \iota_{\gamma} \varepsilon  \,  \psi \right) \right\}, \\[4mm]
\ndelta \td{\mathcal{Q}}^{(T)}_\xi = - \frac{1}{8 \pi G} \int_{\partial \Sigma} \, \delta e^a \swedge \mathcal{L}_{\xi} e_a. \label{dslashQtT}
\end{gather}
Compare these expression with the asymptotic charges corresponding to vacuum Einstein gravity, namely equations \eqref{nH} and \eqref{nHSD}. It is clear that $\ndelta \mathcal{Q}^{(T)}_\xi$ coincides with $\ndelta \mathcal{Q}_\xi$ up to contributions from the fermion fields, while it can also be shown that when the torsion vanishes equation \eqref{dslashQtT} is equivalent to \eqref{nHSD}. 

As in section \ref{sec:charges}, the charges associated with the Lorentz transformation are trivial.  We will not repeat the argument here, since the analysis is essentially identical to that of section \ref{subsec:lorentz}. 

\section{Discussion} \label{sec:dis}

In this paper we have presented a Hamiltonian derivation of the dual BMS charges proposed in Refs.\ \cite{dual0, dualex}.  This derivation justifies their interpretation as asymptotic charges.  The main motivation for the extensions of BMS charges proposed in Refs. \cite{fakenews, dual0, dualex} was to understand Newman-Penrose charges \cite{NP} as BMS charges; that is to give an asymptotic symmetry interpretation of these charges.  In Ref.\ \cite{fakenews}, it was found that a generalisation of standard BMS charges contains half of the set of 10 non-linear Newman-Penrose charges, while it was argued in Ref.\ \cite{dualex} that a new set of dual BMS charges would contain the other five Newman-Penrose charges.  Therefore, a consequence of the results of this paper is that we have finally given a full Hamiltonian derivation of Newman-Penrose charges.

The addition of the Holst term to the Palatini action in section \ref{sec:PH} is controlled by an arbitrary parameter $\lambda$.  Setting $\lambda=0$ gives back the Palatini action, while $\lambda = -1$ corresponds \cite{Holst} to Ashtekar variables, which is a reformulation of general relativity as an SU$(2)$ gauge theory \cite{ashvar}.  There are two other independent arguments for why we ought to choose $\lambda = -1$: In Ref.\ \cite{dualex}, it was found that $\lambda=-1$ reproduces the correct combination of Newman-Penrose charges, while in Ref.\ \cite{Godazgar:2019dkh}, an analysis of the gravitational phase space found that the BMS algebra acts in a well-defined manner only if $\lambda =-1$.  As we observed in section \ref{sec:PH}, $\lambda= \pm 1$ is a somewhat singular choice, since in this case the $P$ operator is non-invertible, see equation \eqref{Pinv}.  In fact, these choices correspond to (anti)-self-dual Palatini gravity \cite{Samuel:1987td,Jacobson:1988yy}. In particular, $\lambda=-1$ projects onto the self-dual part of the Riemann curvature 2-form (or equivalently the self-dual part of the spin connection).  This means that the equations of motion are not clearly Einstein's equation.  In order to resolve this apparent problem, we should recall that in adding the Holst term, we have made the theory complex.  Therefore, we require reality conditions in order to reduce the degrees of freedom to that of the real theory.  When $\lambda \neq \pm 1,$ this is simple to do: we simply require that the solutions be real.  However, when $\lambda = \pm 1$, the reality condition that takes one back to Einstein theory is not as clear, although one does exist \cite{Immirzi:1992ar}, so that even in this case we can be confident that we are working with a theory that is equivalent to Einstein's, albeit not obviously so.  We do not have to worry about the details of this issue here, since the invertibility of the $P$ operator is not required when defining charges.  Therefore, our results are valid for the cases where $\lambda= \pm 1$.

This work raises many further interesting questions that we wish to explore in future work.  One important question is how these ideas can be understood in the context of the Barnich-Brandt formalism \cite{BB}.  This is an alternative formalism for the derivation of asymptotic charges that relies solely on the equations of motion, rather than the presymplectic structures as in the covariant phase space formalism.  The justification for such a formalism is that it relies on the only objects in the theory that matter, namely the equations of motion, rather than objects that have many ambiguities.  For standard BMS charges, it agrees with the expression derived from the covariant phase space formalism, see e.g.\ \cite{Compere}.  However, the main message of Ref.\ \cite{letter} and this work is that there is more to be considered beyond the equations of motion, which seems to go against the spirit of the Barnich-Brandt formalism.  Therefore, a question that we look forward to considering in the near future is whether dual charges can be derived from the Barnich-Brandt formalism at all?  And if so, how?  Related questions have been considered previously in Refs.\ \cite{Torre:1994pf, Barnich:1995ap, Barnich:2008ts}.

We have shown how the Wald-Zoupas prescription can be generalised to define the leading order integrable dual charge in section \ref{sec:int}.  The identification of the integrable charge is an important step in the construction of the charge algebra \cite{BarTro}, which we derived here for leading order dual charges, see section \ref{sec:algebra}.  A construction of the charge algebra for subleading charges \cite{fakenews, dualex} remains to be done.  Of course, one can identify integrable charges order by order and, hence, derive the charge algebra order by order.  However, it would be much more satisfactory to have an all order result.  In order to do this, one must first formulate a Wald-Zoupas prescription for subleading charges.  

In section \ref{sec:charges}, we found that the diffeomorphism and Lorentz generators decoupled.  Investigating each in turn, we found that the charges associated with the Lorentz generators is trivial.   Of crucial importance in deriving this result is the determinant condition \eqref{con:det}.  Therefore, the decoupling of diffeomorphisms and Lorentz generators and the triviality of the Lorentz charges seems to be inextricably linked to our definition of asymptotic flatness, which corresponds to that of Bondi and Sachs \cite{bondi, sachs}.  This is not so surprising since the charges will clearly depend on the background and the boundary conditions that we impose.  In light of this, it would be interesting to consider what happens, for example in the Newman-Unti gauge \cite{newun}?  For standard BMS charges in the metric formulation of the Barnich-Brandt formalism, this has been studied previously and it has been found that the charges in the Newman-Unti gauge satisfy the same charge algebra as those in the Bondi-Sachs gauge \cite{Barnich:2011ty}.

\section*{Acknowledgements}

We would like to thanks Gary Gibbons and Chris Pope for discussions. We would like to thank the Mitchell Family Foundation for hospitality at the 2019 Cook's Branch workshop and for continuing support. M.G.\ and M.J.P.\ would like to thank the Max-Planck-Institut f\"ur Gravitationsphysik (Albert-Einstein-Institut), Potsdam and H.G.\ would like to thank Queen Mary University of London for hospitality during the course of this work. H.G.\ is supported by the ERC Advanced Grant “Exceptional Quantum Gravity” (Grant No. 740209).  M.G.\ is supported by a Royal Society University Research Fellowship. M.J.P.\ is supported by an STFC consolidated grant ST/L000415/1, String Theory, Gauge Theory and Duality.

\appendix

\section{The metric and spin connection} \label{app:spin}

For convenience, in this appendix, we list the metric and inverse metric components
\begin{equation}
 g_{\mu \nu} = e_\mu^a e_\nu^b\, \eta_{ab}, \qquad g^{\mu \nu} = e^\mu_a e^\nu_b\, \eta^{ab},
\end{equation}
as well as the spin connection components 
\begin{equation}
  \omega_{\mu \, ab} = e^{\rho}_{[a} e^{\sigma}_{b]} \left( e_{\sigma \, c} \, \partial_{\mu} e_{\rho}^{c} + \partial_{\sigma} g_{\rho \mu} \right).
\end{equation}
For $X^\mu=(u,r,x^I)$, we have
\begin{gather}
 g_{\mu \nu} = \begin{pmatrix}
                -e^{2\beta} F  + r^2 h_{KL} C^K C^L & -e^{2\beta} & - r^2 h_{JK} C^{K} \\
                -e^{2\beta} & 0 & 0 \\
                - r^2 h_{IK} C^{K} & 0 & r^2 h_{IJ} 
               \end{pmatrix}, \\[3mm]
 g^{\mu \nu} = \begin{pmatrix}
                0 & -e^{-2\beta} & 0 \\
                -e^{-2\beta} & e^{-2\beta} F  & - e^{-2\beta} C^{J} \\
                0 & - e^{-2\beta} C^{I} & r^{-2} h^{IJ} 
               \end{pmatrix},              
\end{gather}
where
\begin{equation}
 h_{IJ} = E_I^i E_J^j\, \eta_{ij}, \quad  h^{IJ} = E^I_i E^J_j\, \eta^{ij}.
\end{equation}
The spin connection components are
\begin{align}
 \omega_{01} &= 2 \, \partial_r \beta\ e^0 + \frac12 e^{-2\beta} \partial_r F\ e^1 + E^I_i \left( \frac{1}{r} \partial_I \beta + \frac{r}{2} e^{-2\beta} h_{IJ} \partial_r C^J \right) e^i, \notag \\[3mm]
 \omega_{0i} &= -E^I_i \left( \frac{1}{r} \partial_I \beta - \frac{r}{2} e^{-2\beta} h_{IJ} \partial_r C^J \right) e^1 - \left( \frac{1}{r} \eta_{ij} + E^I_{(i} \partial_{|r|} E_{j)I} \right) e^j, \notag \\[3mm]
 \omega_{1i} &= - E^I_i \left( \frac{1}{r} \partial_I \beta + \frac{r}{2} e^{-2\beta} h_{IJ} \partial_r C^J \right) e^0 - \frac{1}{2r} e^{-2\beta} E^I_i \partial_I F\ e^1 \notag \\[2mm]
 &\hspace{18mm} + e^{-2\beta} \left( \frac{1}{2r} F\,  \eta_{ij} - E^I_{(i} \partial_{|u|} E_{j)I} + \frac12 F\, E^I_{(i} \partial_{|r|} E_{j)I} - E_{(i}^I E_{j) J}\, {}^{(2)}\nabla_{I} C^J \right) e^j, \notag \\[3mm] 
 \omega_{ij} &= E^I_{[i} \partial_{|r|} E_{j]I}\ e^0 + e^{-2\beta} \left( E^I_{[i} \partial_{|u|} E_{j]I} - \frac12 F\, E^I_{[i} \partial_{|r|} E_{j]I} + E_{[i}^I E_{j] J} {}^{(2)}\nabla_{I} C^J +   {}^{(2)}\omega_{J\, ij} C^J \right) e^1 \notag \\[2mm] 
 &\hspace{100mm} + \frac{1}{r} E_k^J\, {}^{(2)}\omega_{J\, ij}\ e^k,
 \label{spincon}
\end{align}
where in the above equations $$ E_{i \, I} \equiv \eta_{ij} E^{j}_{I} = h_{IJ} E_i^J,$$  ${}^{(2)}\nabla_I$ is the metric connection associated with $h_{IJ}$, i.e.
\begin{equation}
 {}^{(2)}\nabla_I\, h_{JK} = 0
\end{equation}
and ${}^{(2)}\omega_{I\, ij}$ is the spin connection associated with the zweibein $E^i_I$ satisfying 
\begin{equation}
  \partial_{[I} E^i_{J]} + {}^{(2)}\omega_{[I}{}^{i}{}_{j} \swedge E_{J]}^j=0.
\end{equation}

\section{Twisting on the 2-sphere} \label{app:twist}

We define a twisting operation on tensors defined on the 2-sphere \cite{dual0,dualex} as follows.  For a symmetric tensor $X_{IJ}$, its twist 
\begin{equation} \label{Xtwist}
\widetilde{X}^{IJ} = X_K{}^{(I} \varepsilon^{J)K}, \qquad \varepsilon_{IJ} =  \begin{pmatrix}  0 & 1 \\ -1 & 0  \end{pmatrix} \textup{det} \hat{E}^i_I, 
\qquad \varepsilon^{IJ} =  \begin{pmatrix}  0 & 1 \\ -1 & 0  \end{pmatrix} \frac{1}{\textup{det} \hat{E}^i_I}.
\end{equation}
If $X_{IJ}$ is, furthermore, trace-free, i.e.\ $\gamma^{IJ} X_{IJ} = 0,$ 
then $X_K{}^{[I} \varepsilon^{J]K} = 0$. Therefore, $\widetilde X^{IJ}$ is symmetric without the need for explicit symmetrisation
and we can simply write
\begin{equation} \label{twist}
 \widetilde{X}^{IJ} = X_K{}^{I} \varepsilon^{JK}.
\end{equation}
Moreover, we define the twist of a vector $Y^I$ to be
\begin{equation}
 \widetilde{Y}^I = \varepsilon^{IJ} Y_{J}.
\end{equation}

\section{Derivation of the leading dual charge algebra} \label{app:algebra}

In this appendix, we compute the charge algebra given in equations \eqref{Alg:1} and \eqref{Alg:2}.  We begin by considering 
\begin{align}
 \{ \widetilde{\mathcal{Q}}^{(int)}_{0\, \xi_1}, \widetilde{\mathcal{Q}}^{(int)}_{0\, \xi_2} \} - \widetilde{\mathcal{Q}}^{(int)}_{0\, [\xi_1, \xi_2]} &= \delta_{\xi_2} \widetilde{\mathcal{Q}}^{(int)}_{0\, \xi_1} + \widetilde{\mathcal{N}}_{0 \, \xi_2}[\delta_{\xi_1} \phi] - \widetilde{\mathcal{Q}}^{(int)}_{0\, [\xi_1, \xi_2]}  \notag\\[3mm]
 &\equiv \frac{1}{16 \pi G} \int_{\partial \Sigma}\, d\Omega\ \widetilde{k}_{\xi_1, \xi_2}.
\end{align}
Substituting the field transformations \eqref{var:C}, \eqref{var:Ct} and \eqref{var:C2} into the relevant expressions given by equation \eqref{dual0:int} and \eqref{dual0:nint} gives a long expression with three types of terms: terms involving the radiative modes $\partial_u C_{IJ}$ or equivalently $\partial_u \widetilde{C}_{IJ}$; terms involving only the generators of conformal transformation on the round sphere $Y$ and, finally, terms involving a combination of $Y$s and $f$s.  We will look at each set of terms in turn, beginning with the terms involving the radiative modes:
\begin{align}
 \widetilde{k}_{\xi_1, \xi_2} =& 
 f_2\, \partial_u \widetilde{C}^{IJ}\, D_I D_J f_1  - f_1\, D_I D_J \left[ f_2 \partial_u \widetilde{C}^{IJ} \right] \notag \\[3mm]
 &  + \frac14 Y_1^K \left( f_2\, \partial_u \widetilde{C}^{IJ} D_K C_{IJ} - D_K \left[f_2\, \partial_u \widetilde{C}^{IJ}\right] C_{IJ} \right) 
 - \frac14 \widetilde{Y}^I D_I \left( f_2\, \partial_u C^2 \right) \notag\\[3mm]
 & - \frac12 f_2\, \partial_u \widetilde{C}^{IJ} \left( - \half D_K Y_1^K\, C_{IJ} + Y_1^K D_K C_{IJ} + 2 C_{KI} D_J Y_1^K \right) + \ldots \notag\\[5mm] 
 = & D_I \left( f_2\, \partial_u \widetilde{C}^{IJ}\, D_J f_1 \right)  -  D_I \left( f_1\, D_J \left[f_2 \partial_u \widetilde{C}^{IJ} \right] \right) \notag\\[3mm]
 & \hspace{-1mm} - \frac14 Y_1^K  D_K \left( f_2\, \partial_u \widetilde{C}^{IJ} C_{IJ} \right) + \frac14 D_K Y_1^K \, f_2\, \partial_u \widetilde{C}^{IJ} C_{IJ}
 - f_2 \, C^{K}{}_{J} \partial_u \widetilde{C}^{IJ} \, D_{(K} Y_{1 I)} + \ldots, \notag
\end{align}
where we have used the Schouten identity to rewrite
\begin{equation} \label{Sch1}
 - \frac14 \widetilde{Y}^I D_I \left( f_2\, \partial_u C^2 \right) = \frac12 Y_{1 I} D^{K} \left( f_2 \, C_{KJ} \partial_u \widetilde{C}^{IJ} \right)
 -\frac12 Y_1^K D_{I} \left( f_2 \, C_{KJ} \partial_u \widetilde{C}^{IJ} \right).
\end{equation}
Furthermore, we make frequent use above and in what follows of the property that for arbitrary covariant operators $\mathcal O_1$ and $\mathcal O_2$
\begin{equation}
 \mathcal O_1 C_{IK}\, \mathcal O_2 \widetilde{C}^{JK} = - \mathcal O_1 \widetilde{C}_{IK} \, \mathcal O_2 C^{JK},
\end{equation}
which can be proved simply from definition \eqref{twist}.
Now using equation \eqref{Y:CKV}, we find that the terms involving the radiative modes can be grouped into total derivative terms, which can safely be discarded
\begin{align}
 \widetilde{k}_{\xi_1, \xi_2} = 
 D_I \left( f_2\, \partial_u \widetilde{C}^{IJ}\, D_J f_1 \right)  -  D_I \left( f_1\, D_J \left[f_2 \partial_u \widetilde{C}^{IJ} \right] \right) 
 - \frac14  D_K \left( f_2\, Y_1^K\, \partial_u \widetilde{C}^{IJ} C_{IJ} \right) + \ldots.
\end{align}
Next, we consider terms involving solely the $Y$ generators:
\begin{align}
\widetilde{k}_{\xi_1, \xi_2} =& \frac14 Y_1^K \left\{ \left( - \half D_L Y_2^L\, \widetilde{C}^{IJ} + Y_2^L D_L \widetilde{C}^{IJ} + 2 \widetilde{C}_{L}{}^I D^J Y_2^L \right) D_K C_{IJ} \right. \notag\\[3mm]
 & \hspace{40mm} \left. - D_K \left( - \half D_L Y_2^L\, \widetilde{C}^{IJ} + Y_2^L D_L \widetilde{C}^{IJ} + 2 \widetilde{C}_{L}{}^I D^J Y_2^L \right) C_{IJ} \right\} \notag \\[3mm]
 & - \frac14 \widetilde{Y}_1^K D_K D_L \left(C^2 Y_2^L \right) - \frac14 \left( Y_1^L D_L Y_2^K - Y_2^L D_L Y_1^K  \right) \widetilde{C}^{IJ} \, D_K C_{IJ} \notag\\[3mm]
 &+ \frac14 \left( Y_1^L D_L \widetilde{Y}_2^K - Y_2^L D_L \widetilde{Y}_1^K  \right) D_K C^2 + \ldots \notag\\[5mm]
 =& \frac12 Y_1^K \left( - \half D_L Y_2^L\, \widetilde{C}^{IJ} + Y_2^L D_L \widetilde{C}^{IJ} + 2 \widetilde{C}_{L}{}^I D^J Y_2^L \right) D_K C_{IJ} \notag\\[3mm]
  & \hspace{-1.6mm} + \frac14 D_K Y_1^K \left( - \half D_L Y_2^L\, \widetilde{C}^{IJ} + Y_2^L D_L \widetilde{C}^{IJ} + 2 \widetilde{C}_{L}{}^I D^J Y_2^L \right) C_{IJ} \notag\\[3mm]
  & \hspace{-1.6mm} - \frac14 Y_{1}^K D_I D_L \left( Y_2^L C_{JK} \widetilde{C}^{IJ} \right) + \frac14 Y_{1 I} D^K D_L \left( Y_2^L C_{JK} \widetilde{C}^{IJ} \right) \notag\\[3mm]
  & \hspace{-1.6mm} - \frac14 \left( Y_1^L D_L Y_2^K - Y_2^L D_L Y_1^K  \right) \widetilde{C}^{IJ} \, D_K C_{IJ}
 + \frac14 \left( Y_1^L D_L \widetilde{Y}_2^K - Y_2^L D_L \widetilde{Y}_1^K  \right) D_K C^2 + \ldots \notag\\[5mm]
 =& \frac14 D_L \left( Y_1^L Y_2^K D_K C_{IJ} \right) \widetilde{C}^{IJ} - \frac14 Y_2^K D_L Y_1^L\, D_K C_{IJ}\, \widetilde{C}^{IJ} - \frac14  Y_1^L D_L Y_2^K  \widetilde{C}^{IJ} \, D_K C_{IJ} \notag\\[3mm]
 &- \frac12 D_L \left( Y_1^L C_{JK} \widetilde{C}^{IJ} \right) D_I Y_2^K  - \frac14  Y_2^L D_L \widetilde{Y}_1^K D_K C^2 - (1\leftrightarrow2) +\ldots \notag\\[5mm]
  =& \frac14 Y_1^K Y_2^L \, \widetilde{C}^{IJ}\, D_{[K} D_{L]} C_{IJ} + \frac14 C^2 \varepsilon^{IK} Y_2^L \, D_{I} D_{K}  Y_{1 L} - (1\leftrightarrow2) +\ldots, \label{alg:2}
\end{align}
where we have freely integrated by parts and ignored total derivative terms and made free use of Schouten identities to derive equations of the form \eqref{Sch1} and
\begin{equation}
 \widetilde{C}^{IK} C^J{}_{K} = \frac12\, C^2\, \varepsilon^{IJ}. 
\end{equation}
Using the definition of the Riemann tensor
\begin{equation} \label{riemann}
 (D_I D_J - D_J D_I) V^K = R_{IJ}{}^{K}{}_L V^L
\end{equation}
in both of the terms in \eqref{alg:2} gives equal and opposite terms that cancel against one another.  Therefore, the expression of interest reduces to the final set of terms involving a combination of $f$s and $Y$s:
\begin{align}
 \widetilde{k}_{\xi_1, \xi_2} =&  2 f_1\, \varepsilon^{IK}\, D_{(I} D_{J)} D_K D^J f_2 \notag\\[3mm]
  & - f_1 \, D_I D_J \left( - \half D_L Y_2^L\, \widetilde{C}^{IJ} + Y_2^L D_L \widetilde{C}^{IJ} + 2 \widetilde{C}_{L}{}^I D^J Y_2^L \right) \notag\\[3mm]
  & + \frac12 Y_1^K \left( D_K \widetilde{C}^{IJ} \, D_I D_J f_2  -  \widetilde{C}^{IJ} \, D_K D_I D_J f_2  \right) + \widetilde{Y}_1^K D_K \left( C^{IJ} D_I D_J f_2 \right) \notag\\[3mm]
  & + \left( Y_1^K D_K f_2 - Y_2^K D_K f_1 - \half f_2\, D_K Y_1^K + \half f_1\, D_K Y_2^K \right) D_I D_J \widetilde{C}^{IJ}.
  \label{alg:3}
\end{align}
Using equation \eqref{riemann}, as well as the fact that 
\begin{equation} \label{ries2}
 R_{IJKL} = \gamma_{IK} \gamma_{JL} - \gamma_{IL} \gamma_{JK},
\end{equation}
it is fairly simple to show that the first term on the right hand side of equation \eqref{alg:3} vanishes.  Simplifying the remaining terms by integrating by parts and using Schouten identities as before gives
\begin{align} \label{alg:4}
 \widetilde{k}_{\xi_1, \xi_2} = \widetilde{C}^{IJ} \left\{ \frac12 f_1 \, D_I D_J\, D_K Y_2^K + 2 Y_2 D_{[K} D_{I]} D_J f_1 -2 D_J D_{[I} Y_2^K D_{K]} f_1 - (1\leftrightarrow2)\right\}.
\end{align}
Consider the third term
\begin{align}
 - 2 \widetilde{C}^{IJ} D_J D_{[I} Y_2^K D_{K]} f_1 &= - \widetilde{C}^{IJ} \varepsilon_{IK} \varepsilon^{PQ} D_J D_P Y_2^K \, D_Q f_1 \notag\\[3mm]
                                                    &= - \widetilde{C}^{IJ} \varepsilon_{IK} \varepsilon^{PQ} (D_P D_J Y_2^K \, D_Q f_1 - D_P D_Q Y_2^K\, D_J f_1) \notag\\[3mm]
                                                    &= - \widetilde{C}^{IJ} \gamma_{IL} Y_2^L \, D_J f_1,
\end{align}
where in the second equality we have used a Schouten identity and in the third line we have used equation \eqref{Y:CKV} and the fact that $C_{IJ}$ is symmetric and tracefree to show that the first term in the second line vanishes, while we have used equations \eqref{riemann} and \eqref{ries2} to simplify the second term in the second line.  Using equations \eqref{riemann} and \eqref{ries2} to simplify the second term in equation \eqref{alg:4}, we find that $\widetilde{k}_{\xi_1, \xi_2}$ simplifies to 
\begin{equation}
 \widetilde{k}_{\xi_1, \xi_2} = \frac12 \widetilde{C}^{IJ} \left( f_1 \, D_I D_J\, D_K Y_2^K - f_2 \, D_I D_J\, D_K Y_1^K \right).
\end{equation}
This establishes the leading dual charge algebra given by equations \eqref{Alg:1} and \eqref{Alg:2}.
 
\bibliographystyle{utphys}
\bibliography{NP}
\end{document}